\documentclass[times,twocolumn,final]{elsarticle}

\usepackage{cnf}

\usepackage{framed,multirow}

\usepackage{amssymb}
\usepackage{latexsym}
\usepackage{caption}
\usepackage{amsmath}
\usepackage{graphicx}
\usepackage{times}
\usepackage{subfigure}
\usepackage{mathrsfs}
\usepackage{tabularx}
\usepackage{booktabs}
\usepackage{url}
\usepackage{xcolor}
\definecolor{newcolor}{rgb}{.8,.349,.1}

\usepackage{hyperref}

\journal{Combustion and Flame}

\begin{document}

\verso{H. Lu et al.}

\begin{frontmatter}

\title{Filtered turbulent flame model with wrinkling correction on chemical source for non-premixed combustion simulation}

\author[1]{Haoyu \snm{Lu}}

    
\author[1]{Junyi \snm{He}}
 
\author[1]{Lipo \snm{Wang}\corref{cor1}} 
\cortext[cor1]{Corresponding author: Global College, Shanghai Jiao Tong University, 800 Dongchuan RD., Shanghai, 200240, China}
\emailauthor{lipo.wang@sjtu.edu.cn}{L. Wang}

\address[1]{Global College, Shanghai Jiao Tong University, 800 Dongchuan RD., Shanghai, 200240, China}

\begin{abstract}
One of the most critical challenges in turbulent combustion modeling is the chemical source closure. In the recently developed filtered turbulent flame model (FTFM), a one-to-one correspondence between filtered scalar quantities and filtered chemical sources can be constructed by inversely solving the filtered flame equations, without the use of conventional presumed probability density functions (PDFs). However, the turbulence induced flame wrinkling, and thus the enhancement of the chemical source, has not been explicitly considered at the resolved scale. In the present study, the wrinkling effect is analytically quantified in a counterfow flame setup, from which FTFM is then further updated by incorporating such a physics grounded stretching correction on the chemical source. The satisfactory accuracy and robustness of the present model are justified from case tests of the non-premixed Sydney swirl flame and the Delft III flame.
\end{abstract}

\begin{keyword}
\KWD Filtered turbulent flame model\sep Flame wrinkling\sep Filtered chemical source stretching\sep Non-premixed turbulent combustion\sep Large eddy simulation
\end{keyword}

\end{frontmatter}


\section*{Novelty and significance statement}
Regarding the filtered turbulent flame model for non-premixed combustion simulations, there are two prominent advantages. One is that the tabulation entry parameters are numerically available; the other is the closure between the chemical source and species by inverse mapping. Further improvement is achieved to analytically quantify the flame wrinkling effect on the chemical source in a counterflow flame setup. Overall, the present framework offers an alternative insight in turbulent combustion modeling.

\section{Introduction\label{sec:introduction}} 
\addvspace{10pt}
In turbulent combustion, because of the broad spectra of both the temporal and spatial scales, the trade-off between the numerical cost and efficiency makes the large eddy simulation (LES) practically preferable. Within the LES framework, for reacting flows at the subgrid-scale level, the major modeling challenge comes from the turbulence-chemistry interaction, i.e. the relation between the filtered chemical sources and filtered scalar quantities, including temperature and species concentrations. 

There are numerous efforts devoted for this challenging issue. For instance, in the conditional moment closure method~\cite{CMC}, the conditional mean quantities, higher moments as well, are modeled conditional on given values of the mixture fraction or reaction progress variable, while in the conditional source-term estimation (CSE)~\cite{cse-review},  the chemical sources are conditionally calculated using the conditional scalars. The flamelet modeling strategy is particularly popular because of the clear physical interpretation and numerical efficiency. Through the flamelet transform, the dimensionality of relevant field variables is largely reduced, meanwhile, the fundamental flame physics is still preserved. Various numerical implementations based on this concept have been developed thereafter~\cite{williams1985combustion,PETERS19881231,peters1984laminar,PIERCE_MOIN_2004}. 

In LES, the focus of the flamelet modeling part is to close the filtered species and chemical sources from the laminar ones, for instance using presumed probability density functions (PDFs), such as the $\beta$-function, which will then inevitably introduce arbitrary modeling input parameters. Auzillon et al.~\cite{FIORINA2010465} proposed to construct the closure directly from the filtered field. In the filtered tabulated chemistry model for LES (F-TACLES), the unclosed terms are obtained by filtering the one-dimensional premixed laminar flame solutions. In the analysis of the filtered species equation~\cite{WANG2017259}, a differential relation between filtered species mass fractions and the filtered mixture fraction at the resolved scale can be revealed after a flamelet-like transform. The numerical results demonstrated clear scaling properties of the filtered flame structure with respect to the filtering length. In addition, the normal direction of filtered flame fronts was shown to align well with the gradients of species mass fractions, which allows further simplification of the analysis. Based upon this theoretical foundation, the filtered turbulent flame model (FTFM) is proposed~\cite{zhang2023ftfm}. Unlike F-TACLES, in which the chemical sources are still modeled from the filtered quantities, in FTFM the filtered quantities are determined from the filtered chemical sources. In this sense, there is no need to introduce the modeled PDFs and thus no closure arbitrariness. Mathematically, if the filtering length shrinks to zero, the FTFM and flamelet formulations are in nature compatible. Overall, numerical results from FTFM are acceptable, in spite of discrepancies in species predictions~\cite{zhang2023ftfm}. In a recent update of FTFM-C framework~\cite{He13082025}, the scalar dissipation rate $\chi_{t}$, one of the tabulation entry parameters, is replaced by the filtered progress variable $\widetilde{C}$, resulting in much improved numerical stability and accuracy. 

Despite these progresses, in the FTFM framework, there remains an important physical mechanism, which is not yet sufficiently addressed. Since in turbulent reacting flows, the flame surface wrinkling, and stretching as well, will largely enhance the effective flame surface area and the overall consumption rate~\cite{Peters_2000}, although the internal flame structure is well preserved. In premixed turbulent combustion, the unresolved flame surface area can be quantified by its density (FSD) transport model equation~\cite{Trouvé_Poinsot_1994}. Subsequently, various flame wrinkling models in LES, including the efficiency function approach for thickened flame (TF) model~\cite{Colin2000}, power-law flame wrinkling model~\cite{CHARLETTE2002159} and fractal flame model~\cite{FUREBY2005593}, have been proposed to account for the unresolved flame wrinkling. Similarly, in non-premixed turbulent combustion, direct numerical simulations~\cite{VK1998} also suggest significant flame wrinkling. In the existing versions of FTFM~\cite{zhang2023ftfm,He13082025}, such flame wrinkling effect was mainly incorporated from the dimensional analysis.

In the present work, we focus on this important flame wrinkling issue in FTFM, which will be re-evaluated upon a physics guaranteed basis. Therefore, it is naturally expected that the physical consistency of filtered chemical source closure will be improved. The updated FTFM with reformulated chemical source, is now referred to as filtered turbulent flame model with source stretching (FTFM-ST). The rest of the work is organized as follows. First, a brief overview of the filtered flame concept and the model construction is present. Second, the theoretical analysis for flame wrinkling correction and tabulation setup are elaborated. After the simulation results for the Sydney swirl flame and the Delft III flame, in comparison with experimental data and other model predictions, the capability and application potential of FTFM-ST will then be summarized.

\section{Model construction} \addvspace{10pt}
\subsection{Flamelet transform of the filtered scalar equations}\addvspace{10pt}
First, the model foundation is briefly described. Similar to the laminar flamelet transform, the filtered flame equation, under the assumptions of the unity turbulent Lewis number and negligible tangential derivatives of the filtered quantities, is written as~\cite{WANG2017259}
\begin{equation}
    \overline{\rho}\frac{\partial{\widetilde{Y}_i}}{\partial \tau}=\frac{1}{2}\overline{\rho}\chi_t \frac{\partial^2 \widetilde{Y}_i}{\partial \widetilde{Z}^2} + \overline{\omega_i(Y_1,Y_2,...,T)},
\label{GE}\end{equation}
where $\overline{\rho}$ is the filtered density, $\widetilde{Z},\widetilde{Y}_i,\text{and }\widetilde{T}$ represent filtered Favre-averaged mixture fraction, species concentration and temperature, respectively. The turbulent scalar dissipation rate $\chi_t$ is defined as
\begin{equation}
    \chi_t = 2\mathfrak{D}_t \left(\frac{\partial \widetilde{Z}}{\partial n}\right)^2.
\end{equation}
Here $n$ is the normal directional coordinate of the filtered flame surface, the effective diffusivity $\mathfrak{D}_t=\widetilde{D}+D_t$, where $\widetilde{D}$ is the Favre-averaged laminar diffusivity and $D_t$ is the subfilter diffusivity that lumps the unresolved turbulent mixing effect. The meaningfulness of Eq.~\eqref{GE} is that it shows the functional dependence of the filtered quantities on the filtered density $\overline{\rho}$, filtered mixture fraction $\widetilde{Z}$, turbulent scalar dissipation rate $\chi_t$ and the filtered chemical source $\overline{\omega_i}$. Interestingly, except for the filtered chemical sources, other quantities ($\overline{\rho}$, $\widetilde{Z}$ and $\chi_t$) can be directly obtained from the resolved fields, thereby eliminating their need for assumed PDFs in subgrid closure.


\subsection{Flame wrinkling factor}\addvspace{10pt}\label{Section:Fw}

\begin{figure*}[t!]
\centering
\includegraphics[width=0.6\textwidth]{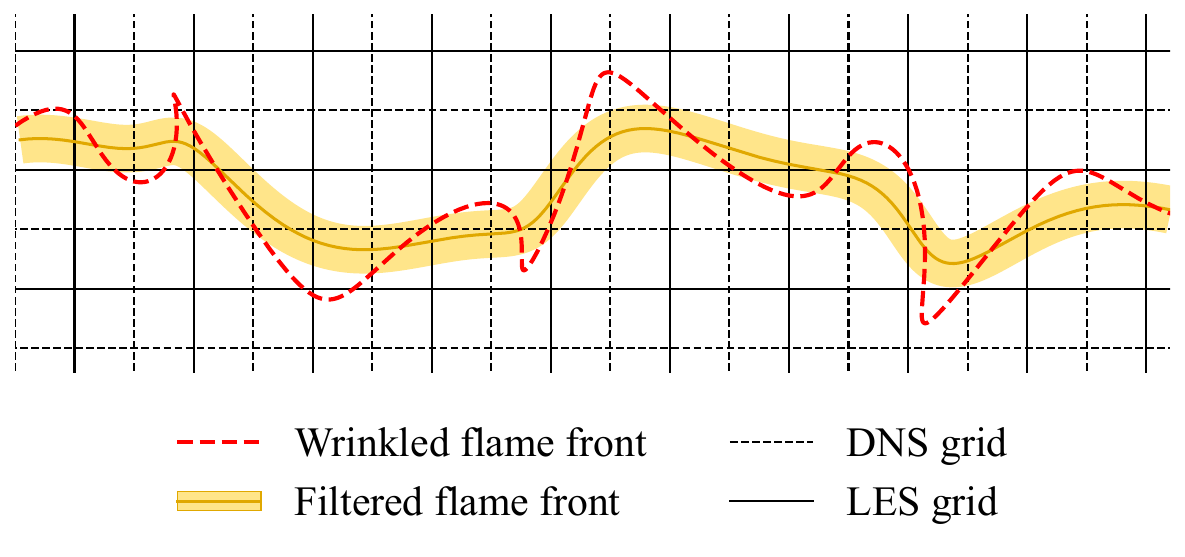}
\caption{\footnotesize Schematic of turbulence-induced flame stretching and filtering at the LES scale. The red dashed line denotes the fully resolved wrinkled flame front at the DNS scale level (dashed grid); the yellow band represents the flame zone after filtering at the LES scale (solid grid), with enlarged thickness because of the turbulence-induced stretching correction; the dark-yellow centerline inside the flame zone denotes the corresponding filtered flame-front location.}
\label{flamegrid}
\end{figure*}

As the schematic in Fig.~\ref{flamegrid}, the flame front becomes wrinkled by turbulent disturbance, which increases the flame surface area and amplifies the total consumption rate within the flame zone. Statistically, both the turbulent flame zone thickness and the peak of the chemical reaction rate are turbulence intensity dependent. Such a turbulence effect is quantified by a chemical source associated wrinkling factor, which was determined from a dimensional argument in the previous formulation~\cite{zhang2023ftfm,He13082025}. In the present work, this wrinkling correction will be revisited from physics guaranteed relations.   


Along the flame normal, consider a quasi-one-dimensional laminar counterflow with the axial velocity profile $u=-ax$ and fuel mass fraction $Y_{f}$ at the fuel inlet. The fuel (or other species) mass flux across the flame is consumed by the chemical reaction, which leads to 
\begin{equation}
\int_{-\delta/2}^{\delta/2} \omega(x)dx = \mathcal{C}\delta, \label{eq:laminarRR}
\end{equation}
where $\mathcal{C}=\dfrac{1}{2}a\rho Y_{f}$ and $\delta$ denotes the flame thickness. In turbulence at the resolved scale, Eq.~\eqref{eq:laminarRR} holds as well in a statistical average sense. The effective flame thickness increases to $\delta^*$ and the local equivalent turbulent diffusivity is $D_t$. Because of the scaling relation $\delta \propto \sqrt{D/a}$, a thickness ratio $\mathcal{A}$ is then
\begin{equation}
\mathcal{A}=\frac{\delta^*}{\delta} = \sqrt{\frac{D+D_t}{D}}.\label{eq:stretchfactor}
\end{equation}
Meanwhile, the species consumption relation satisfies
\begin{equation}
\int_{-\delta^*/2}^{\delta^*/2} {\omega}^*(x)dx = \mathcal{C}\delta^*,
\end{equation}
where ${\omega}^*$ represents the statistical mean chemical source with turbulence. Note that the coefficient $\mathcal{C}$ remains the same as in Eq.~\eqref{eq:laminarRR} since turbulence does not change the imposed inlet conditions. 

Introducing coordinate stretching $\zeta=x/\mathcal{A}$ and using Eq.~\eqref{eq:stretchfactor}, we obtain
\begin{equation}
    \int_{-\delta/2}^{\delta/2}{\omega}^*(\mathcal{A} \zeta)d\zeta=\mathcal{C} \delta. \label{eq:turbRRtrans}
\end{equation}
If the profiles of $\omega(x)$ and ${\omega}^*(x)$ are geometrically similar, comparison between Eq.~\eqref{eq:laminarRR} and Eq.~\eqref{eq:turbRRtrans} yields
\begin{equation}
    {\omega}^*(\mathcal{A} \zeta)=\omega(\zeta), \ \text{where}\ \zeta\in[-\frac{\delta}{2},\frac{\delta}{2}],
\end{equation}
i.e. in original $x$
\begin{equation}
    {\omega}^*(x)=\omega\left(\frac{x}{\mathcal{A}}\right),\ \text{where}\ x\in[-\frac{\mathcal{A}\delta}{2},\frac{\mathcal{A}\delta}{2}]. \label{eq:stretch1}
\end{equation}

An important conclusion here is that the way turbulence can enhance the chemical source is exerted by a spatial stretching of the reaction zone, i.e. ${\omega}_i(x)\xrightarrow{}{\omega}^*_i\left(x/\mathcal{A}\right)$, while the source peak does not change. The total consumption rate  $\Omega_{i}^{*}$ is then
\begin{equation}
\scalebox{1.0}{$
\begin{aligned}
\Omega_{i}^{*}
&= \int_{-\infty}^{\infty}{\omega_i}^*(x)dx= \int_{-\infty}^{\infty}\omega_i\!\left(\frac{x}{\mathcal{A}}\right)\,dx
\\&
= \mathcal{A} \int_{-\infty}^{\infty}\omega_i\!\left(\frac{x}{\mathcal{A}}\right)
\,d\!\left(\frac{x}{\mathcal{A}}\right)= \mathcal{A}\,\Omega_{i}.\label{eq:stretch2}
\end{aligned}
$}
\end{equation}
Thus the total consumption rate is enhanced through the reaction zone stretching factor $\mathcal{A}$, i.e. the index to characterize the chemical source enhancement by turbulent diffusion.

\subsection{Tabulated database}\addvspace{10pt}
The low-dimensional mapping implied in Eq.~\eqref{GE} can be constituted from the solutions in the physical space. In details, the one-dimensional filtered flame equations are~\cite{He13082025}
\begin{align}
&\overline{\rho}\frac{\partial \widetilde{Z}}{\partial t} 
+ \overline{\rho u}\frac{\partial \widetilde{Z}}{\partial x} 
= \frac{\partial}{\partial x}\Big(\overline{\rho}\mathfrak{D}_t \frac{\partial \widetilde{Z}}{\partial x}\Big),\label{eq.1}\\[1ex]
&\overline{\rho}\frac{\partial \widetilde{Y}_i}{\partial t} 
+ \overline{\rho u}\frac{\partial \widetilde{Y}_i}{\partial x} 
= \frac{\partial}{\partial x}\Big(\overline{\rho}\mathfrak{D}_t \frac{\partial \widetilde{Y}_i}{\partial x}\Big) + \overline{\omega_i^*},\label{eq.2}\\[1ex]
&\overline{\rho C_p}\frac{\partial \widetilde{T}}{\partial t} 
+ \overline{\rho u C_p}\frac{\partial \widetilde{T}}{\partial x} 
= \frac{\partial}{\partial x}\Big(\overline{\rho C_p}\mathfrak{D}_t \frac{\partial \widetilde{T}}{\partial x}\Big) \notag\\
&\quad - \sum_i \overline{j_i \frac{\partial h_i}{\partial x}} 
- \sum_i \overline{\omega_i^* h_i}.\label{eq.3}
\end{align}
In the above equations, $C_p$ is the specific heat capacity, $j_i$ and $h_i$ are the diffusive mass flux and the specific enthalpy of specie $i$. Such one-dimensional formulation is consistent with Eq.~\eqref{GE} in the transformed space. Here, the filtered chemical source term $\overline{\omega_i^*}$ is a function of filtered scalars $\widetilde{Y_{1}},\widetilde{Y_{2}},...,\widetilde{T}$.


\begin{figure*}[htbp!]
\centering
\includegraphics[width=0.97\linewidth]{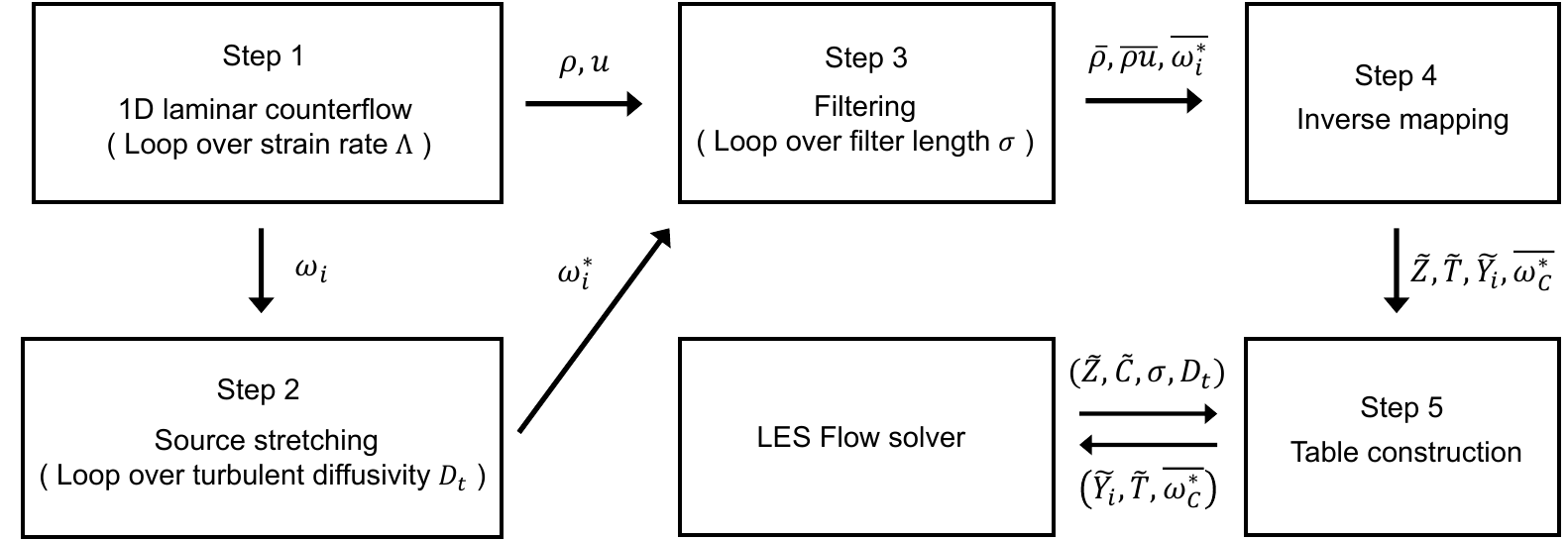}
\caption{\footnotesize Tabulation and numerical procedure to implement FTFM-ST. In step 2 the chemical source stretching operation follows Eq.~\eqref{eq:stretch1}. In step 4 the inverse mapping operation follows Eq.~\eqref{eq.1}-\eqref{eq.3}.}
\label{tabulation}
\end{figure*}

To establish a tabulated database that preserves the one-to-one correspondence between filtered sources and filtered scalars, it is important to utilize the inverse mapping, where the filtered thermochemical scalars are calculated by solving Eqs.~\eqref{eq.1}--\eqref{eq.3} with prescribed filtered chemical sources. For the proposed FTFM-ST model, the tabulation procedure is elaborated as follows.

1. Calculate the one-dimensional laminar counterflow diffusion flame (encoded in Cantera v2.5.1~\cite{goodwin2018cantera}) and loop over different strain rates $\Lambda$ until extinction to obtain a fully resolved flamelet library at the fine scale.

2. From Eq.~\eqref{eq:stretch1}, stretch the laminar chemical sources of each specie $\omega_i$ calculated from step 1 with the corresponding diffusivity $D$ and prescribed subgrid diffusivity $D_t$ to obtain the turbulence-induced stretched source $\omega_i^*$. The stretching is applied with respect to the stagnation point of the one-dimensional counterflow flame.

3. Filter the obtained laminar results from both step 1 and 2 using the Gaussian filter and loop over different filter lengths $\sigma$ to calculate coefficients $\overline{\rho},\overline{\rho u},\overline{\omega_i^*}$, etc., necessitated for solving Eq.~\eqref{eq.1} to~\eqref{eq.3}. Since the filtering is explicitly applied to the fully resolved laminar flamelet solutions, the resulting filtered flame structure is numerically regarded as the reference LES solution for the subsequent model construction.

4. From the filtered stretched  chemical sources, together with coefficients obtained in step 3, the filtered quantities $\widetilde{Z}, \widetilde{Y}_i$ and $\widetilde{T}$ are solved from the linear governing equations Eq.~\eqref{eq.1} to~\eqref{eq.3} using the SparseLU solver (Eigen~\cite{eigen_sparse_docs}). Inverse mapping in the present context implies that the function $\overline{\omega_i^*}(\widetilde{Y_{1}},\widetilde{Y_{2}},...,\widetilde{T})$ is first determined. Then the filtered scalars are retrieved via the closed linear equation system, i.e. mathematically an inverse mapping from the function $\overline{\omega_i}$ to the one-to-one correspondence variable set $(\widetilde{Y_{1}},\widetilde{Y_{2}},...,\widetilde{T})$. 

5. All the physical space solutions are cast to a four-dimensional look-up table with parameters ($\widetilde{Z},\chi_t,\sigma, D_t$). In a subsequent update~\cite{He13082025} from FTFM to FTFM-C, it is advantageous to replace the $\chi_t$ quantity by the filtered progress variable, which can primarily enhance the model robustness and ability to capture extinction and reignition. Here the filtered progress variable $\widetilde{C}=\widetilde{Y}_{\text{H}_2\text{O}}+\widetilde{Y}_{\text{CO}}+\widetilde{Y}_{\text{CO}_2}$. In the flow solver, the transport equations of $\widetilde{Z}$ and $\widetilde{C}$ are solved from the respective governing equations as:
\begin{equation}
    \begin{aligned}
        \frac{\partial}{\partial t}\left(\bar{\rho} \widetilde{Z}\right)+\nabla \cdot \left(\overline{\rho u}\widetilde{Z}\right) &= \nabla\cdot\left(\bar{\rho}\mathfrak{D}_t\nabla \widetilde{Z}\right),\\
        \frac{\partial}{\partial t}\left(\bar{\rho} \widetilde{C}\right)+\nabla \cdot \left(\overline{\rho u}\widetilde{C}\right) &= \nabla\cdot\left(\bar{\rho}\mathfrak{D}_t\nabla \widetilde{C}\right) + \bar{\rho}\overline{{\omega}_C^*},
    \end{aligned}\label{eq:transZC}
\end{equation}
where the filtered source $\overline{\omega_C^*}$ is directly available from the look-up table. Together with the local mesh size as the filter length $\sigma$ and $D_t$ determined by the subgrid-scale (SGS) model, the flame solution at each grid cell can be retrieved by table look-up in the manifold spanned by ($\widetilde{Z},\widetilde{C},\sigma, D_t$). The above algorithm workflow is summarized in Fig.~\ref{tabulation}.

\subsection{Effect of flame wrinkling correction on flamelet solutions}\label{FWdiscussion}\addvspace{10pt}

  \begin{figure*}[h!]
  \centering\vspace{-1mm}
  \includegraphics[width=0.97\linewidth]{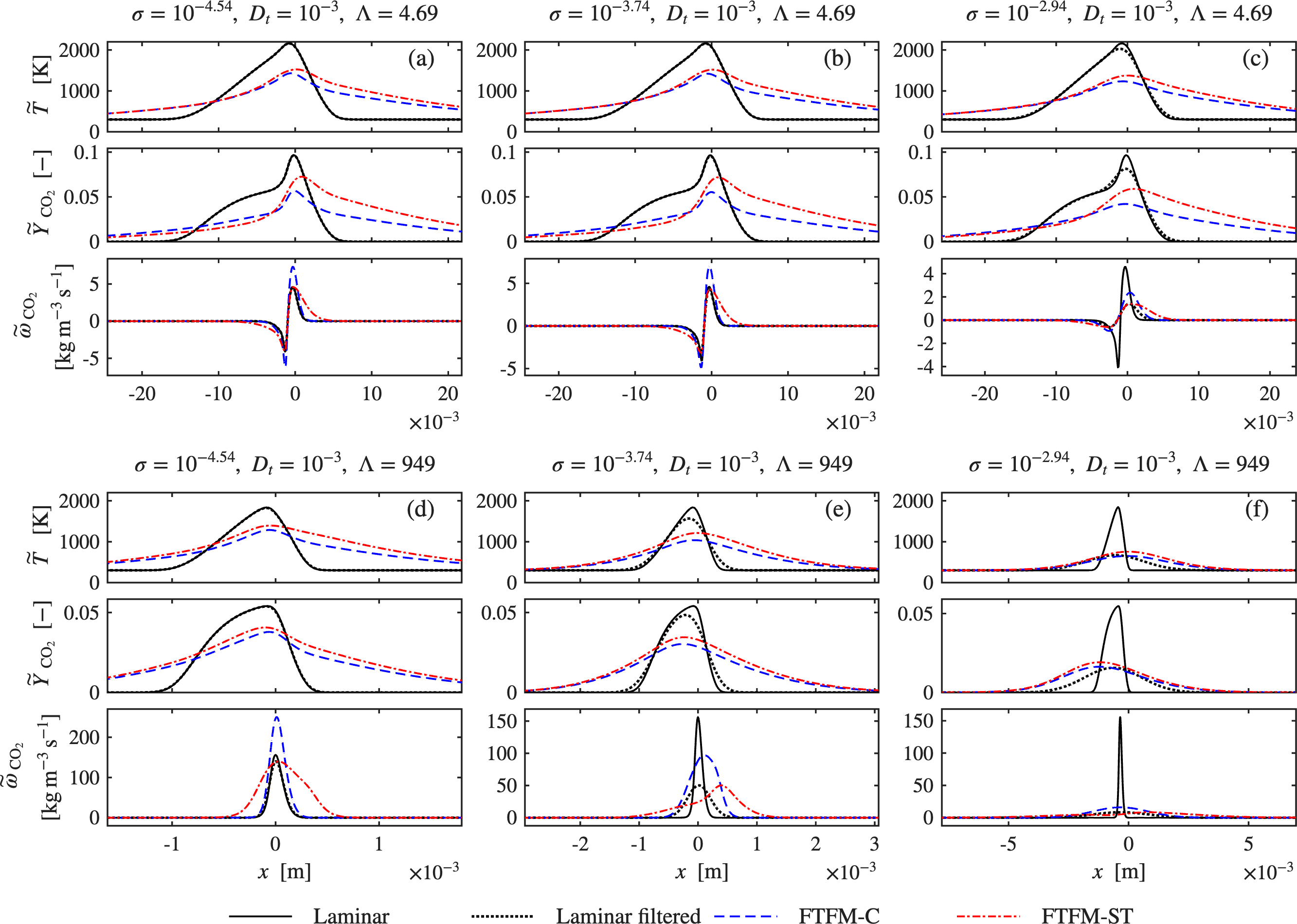}\vspace{-0mm}
  \caption{FTFM-ST flame structure in the physical space at selected states, in comparison with these from laminar, filtered laminar and FTFM-C solutions. The profiles of temperature $\widetilde{T}$, CO$_2$ mass fraction
  $\widetilde{Y}_{\mathrm{CO_2}}$, and CO$_2$ source term
  $\widetilde{\omega}_{\mathrm{CO_2}}$ are shown for different filter width
  $\sigma$ and strain rate $\Lambda$ at a sufficiently large turbulent diffusivity
  $D_t=10^{-3}$.}
  \label{Fig:flameletSol1}\vspace{-1mm}
  \end{figure*}

To clarify the physics and quantitative effect of the stretching correction on the chemical source in FTFM-ST, the one-dimensional flamelet solutions with stretching will be compared with these without stretching, i.e. ${\omega}^*(x)=\mathcal{A}\omega(x)$ as in the FTFM-C formulation~\cite{He13082025}. Since the stretching factor $\mathcal{A}$ approaches unity as $D_t$ decreases, as a consequence, the corrected source profiles and scalar profiles thereafter determined will gradually shrink to the filtered laminar results. Therefore, a relatively large turbulent diffusivity, $D_t=10^{-3}$, is selected to highlight the turbulence influence on the filtered flamelet solutions.

\begin{figure*}[h!]
\centering\vspace{-1mm}
\includegraphics[width=0.95\textwidth]{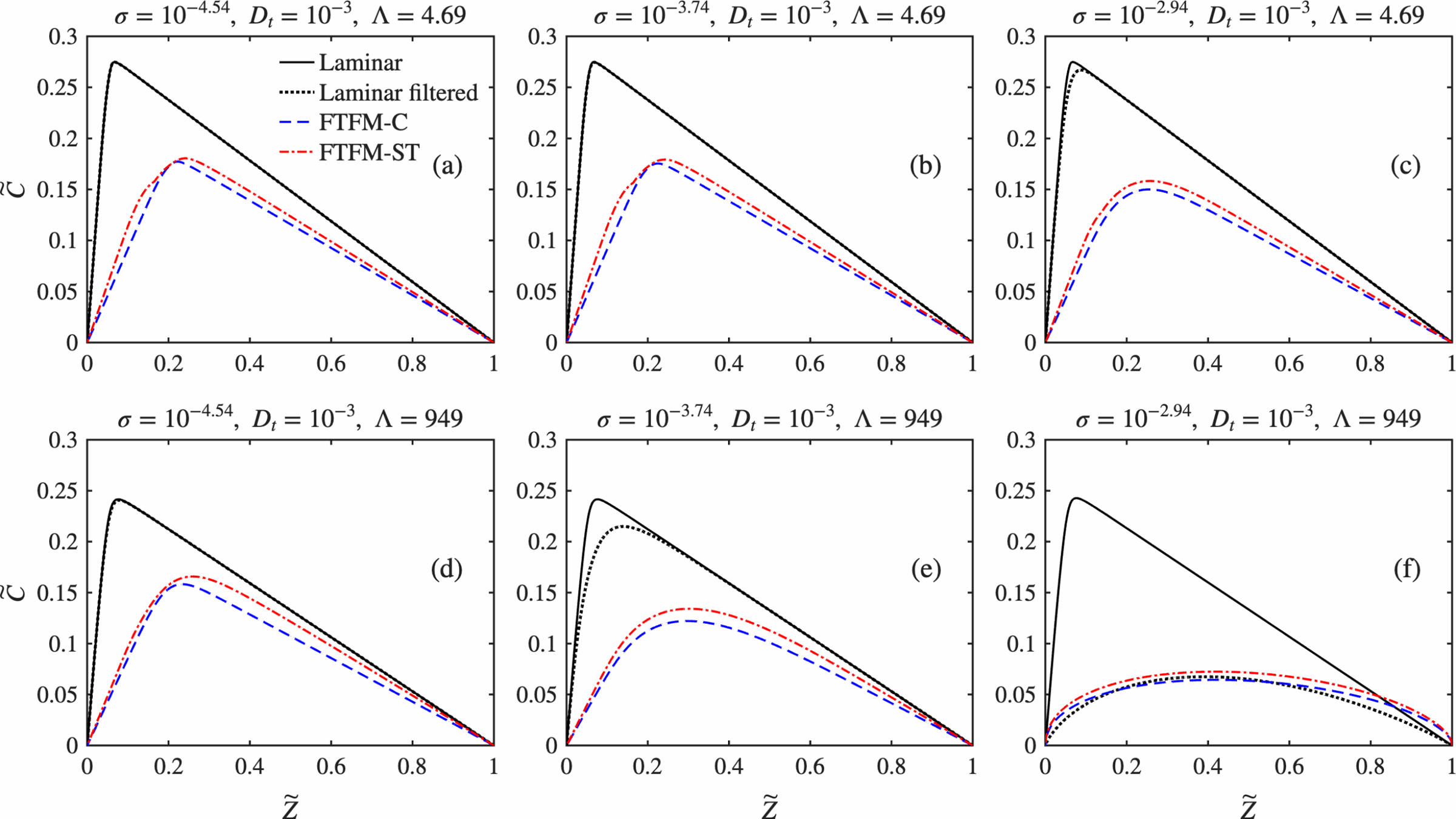}
\caption{Comparison of the selected flamelet solutions in $\widetilde{Z}$--$\widetilde{C}$ space, at the same turbulent diffusivity $D_t=10^{-3}$, filter width $\sigma$ and strain rate $\Lambda$ as in Fig.~\ref{Fig:flameletSol1}.}
\label{Fig:flameletSol2}\vspace{-1mm}
\end{figure*}

Figure~\ref{Fig:flameletSol1} presents the flamelet solutions in physical space, including the laminar solution, the filtered laminar solution, and solutions from FTFM-C and FTFM-ST. Panels (a)-(c) correspond to the low strain rate case at $\Lambda=4.69$, with an increasing filter width $\sigma$ from left to right, while panels (d)-(f) show the same variation of $\sigma$ for the high strain rate case at $\Lambda=949$. For all the selected cases, basically the temperature of FTFM-ST is higher than that of FTFM-C, together with visible changes in the CO$_2$ distribution. Such difference must be from the difference in the filtered source terms between two models. Since the source profile of FTFM-ST is broader, the temperature undergoes a longer ascending process and thus the peak value is greater. The effect of turbulent diffusivity $D_t$ can also be observed in comparison with the filtered laminar solutions. The scalar and temperature profiles inversely determined from their corresponding chemical sources are clearly broader than the filtered laminar ones. At a relatively large strain rate $\Lambda=949$, as shown in panels (d)-(f), this turbulence induced broadening effect becomes weakened with the increasing filter width $\sigma$ from left to right.

Figure~\ref{Fig:flameletSol2} shows the same solutions but mapped in the $\widetilde{Z}$--$\widetilde{C}$ parameter space. With the increase of the filter size, either from (a) to (c) or (d) to (f), the peak values of FTFM-C and FTFM-ST solutions (as well as the filter laminar solution) drop. At the low strain rate ($\Lambda=4.69$) such drop is relatively slight, while at the high strain rate ($\Lambda=949$) the drop is clearly stronger. When $\Lambda$ increases, for instance from (c) to (f), the filtered profiles become more distributed. It is worth noting that for all the cases, the difference between FTFM-C and FTFM-ST solutions keeps moderately visible without clear changing tendency. However, FTFM-C and FTFM-ST solutions are clearly distinct from the filtered laminar ones, as in most cases shown in panels (a)-(e), except in panel (f), which can be interpreted from the relative magnitude of two scales: the flame thickness and the filter size. Physically, the diffusive flame thickness $l_t$ is estimated as $l_t\sim\sqrt{D_t/\Lambda}$. In panel (f), $l_t \sim 10^{-2.98}$, which is comparable to the imposed filter width. Thus the filtering effect is important to change the filtered flame structure. In comparison, the filter size of other cases in panel (a)-(e) are clearly smaller than the flame thickness, the filtered laminar curves are relatively insensitive to filtering effect. In summary, the mapping profiles in Fig.~\ref{Fig:flameletSol2} are dominated not only by the chemical source, but also by the interplay among characteristic scales of the filtered field. 

\subsection{Characteristics of the FTFM-ST manifold}
\begin{figure*}[!h]
\vspace{-5mm}
\centering
\begin{tabular}{cc}
\subfigure[]{\includegraphics[width=0.45\textwidth]{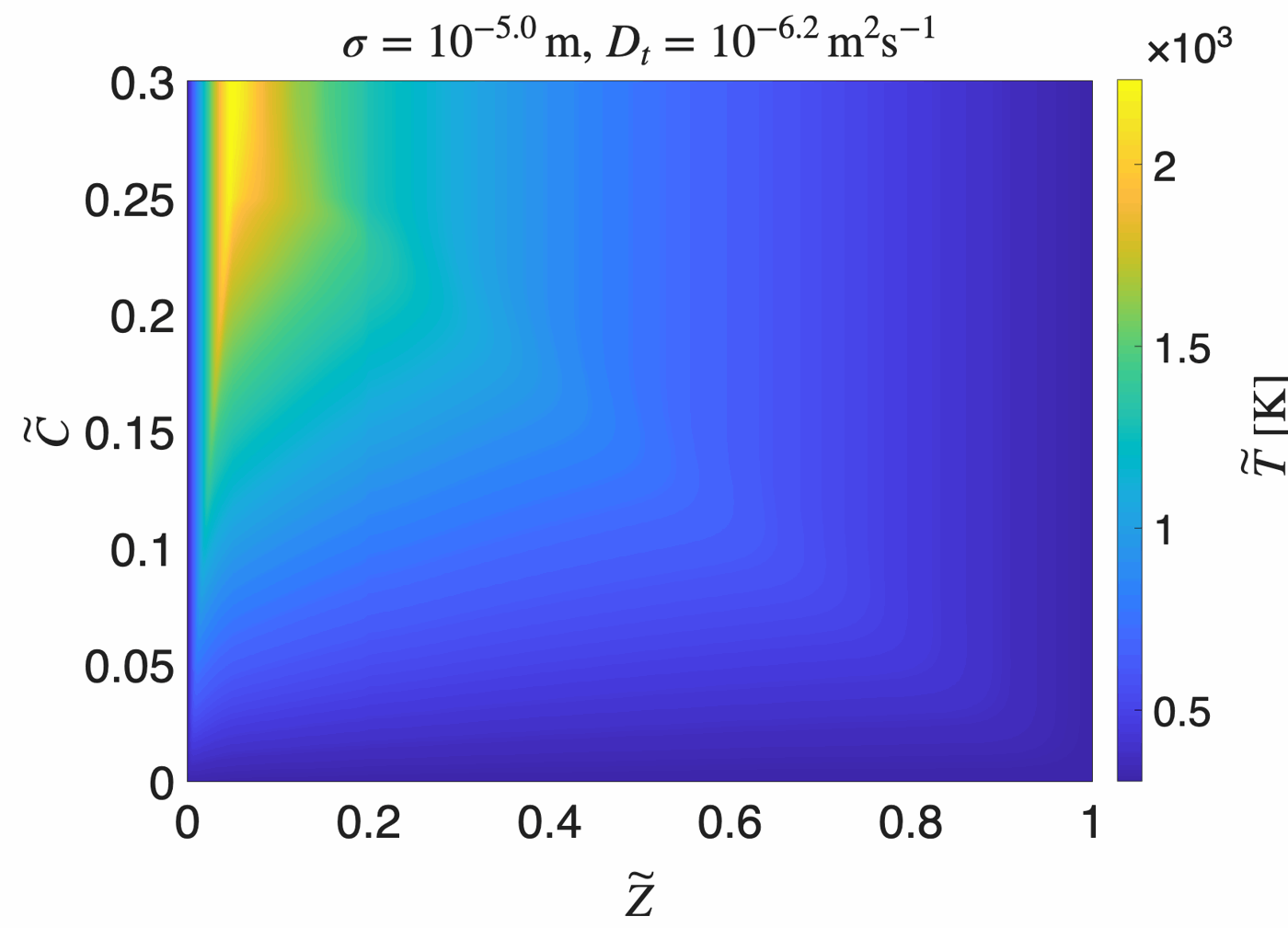}} &
\subfigure[]{\includegraphics[width=0.45\textwidth]{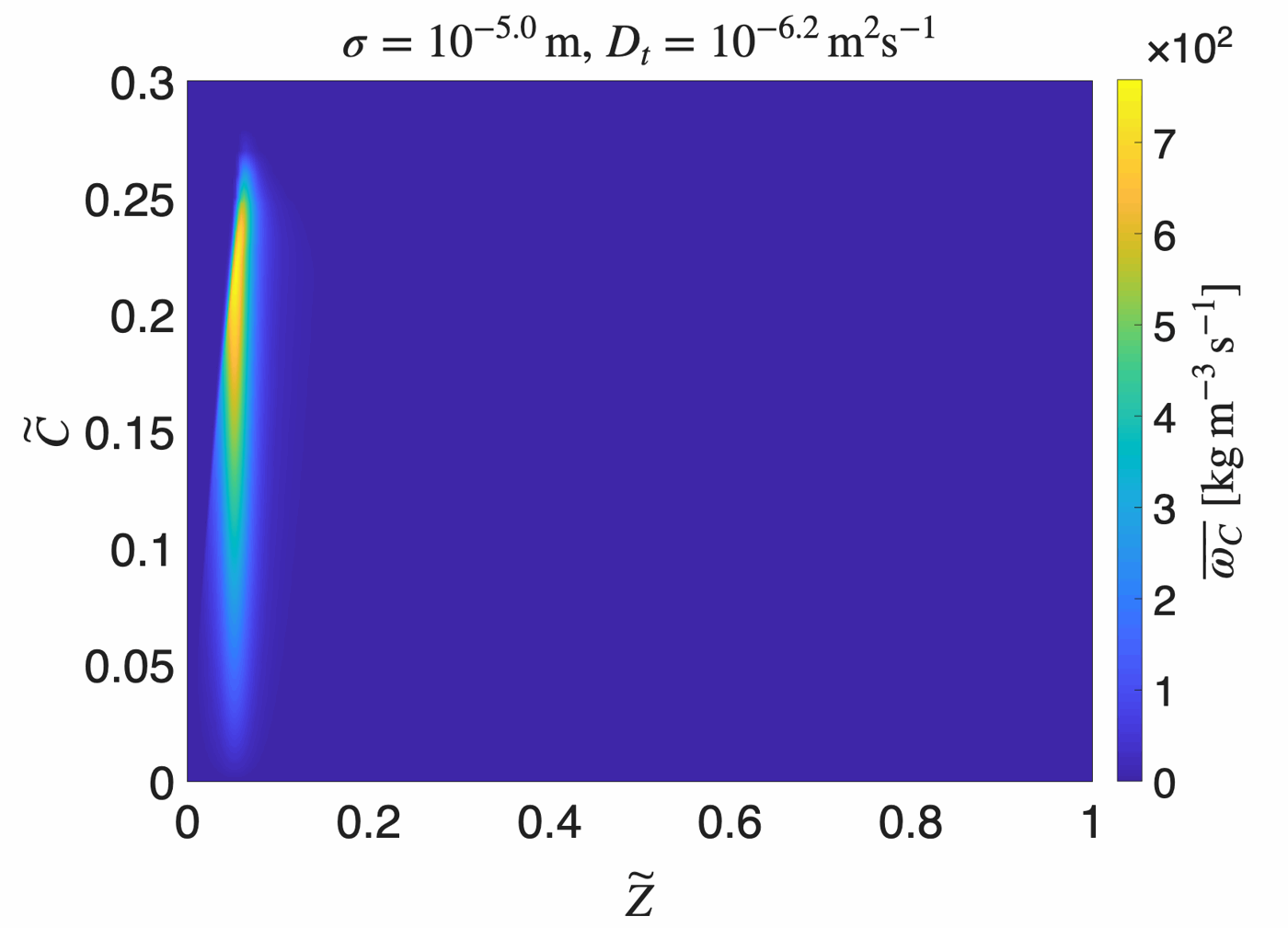}} \vspace{-3mm}\\ \vspace{-3mm}
\subfigure[]{\includegraphics[width=0.45\textwidth]{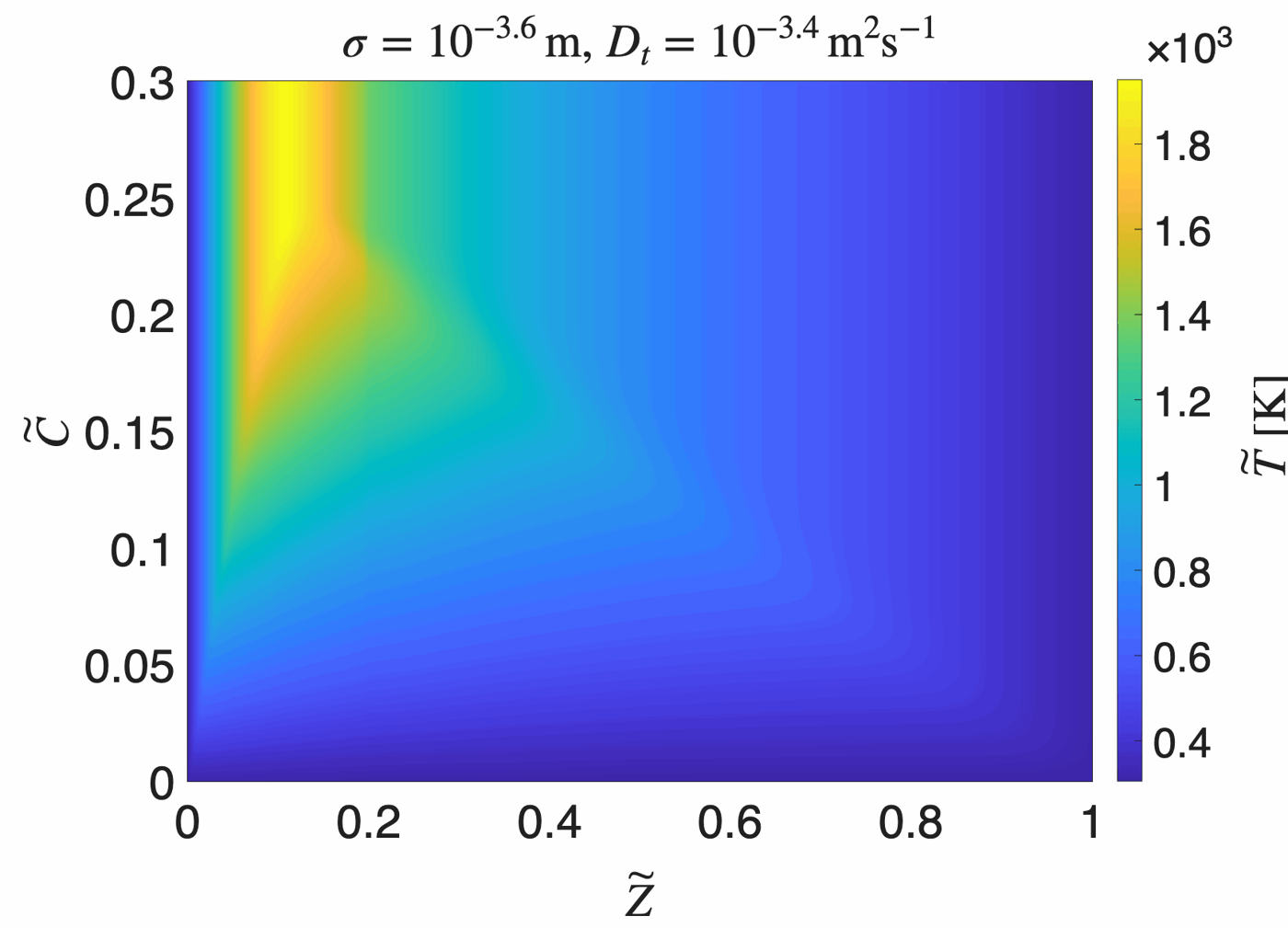}} &
\subfigure[]{\includegraphics[width=0.45\textwidth]{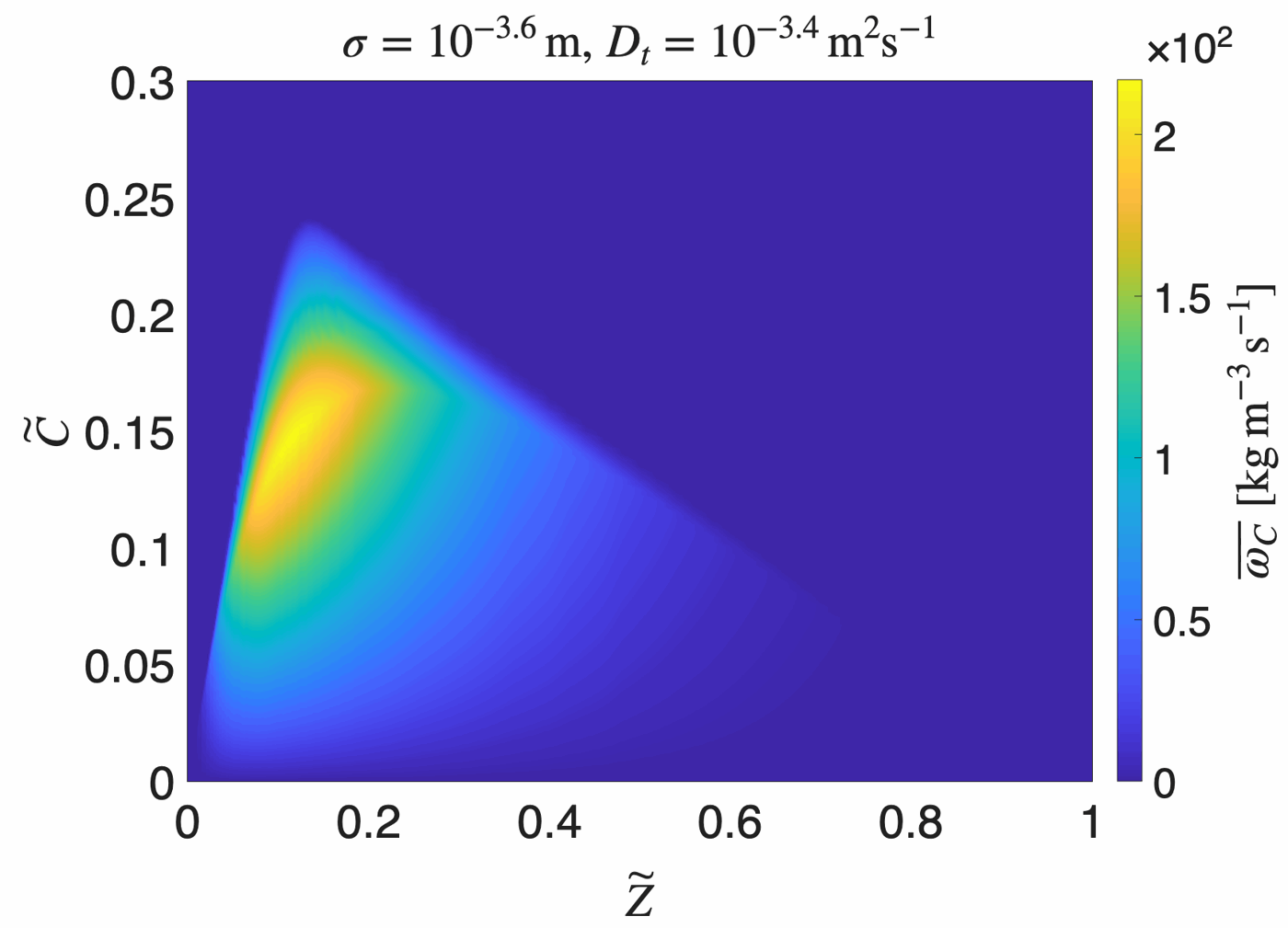}} 
\end{tabular}
\caption{Contours of the temperature $\widetilde{T}$ and progress variable source term $\overline{\omega_C}$ at different filter length $\sigma$ and subfilter diffusivity $D_t$ in the $\widetilde{Z}-\widetilde{C}$ space: $(a,b)$ at the smallest $\sigma$ and $D_t$ limit, representing a fine-scale laminar case; $(c,d)$ at larger $\sigma$ and $D_t$, representing a coarser grid scale LES case.}
\label{table-solution}
\vspace{-5mm}
\end{figure*}
The structure of the constructed FTFM-ST manifold can be understood from representative thermochemical mapping relations. Figure~\ref{table-solution} visualizes the results obtained from the methane/hydrogen mixture with $1:1$ volumetric ratio. At a smaller filter length $\sigma$ and subfilter diffusivity $D_t$ to represent a fine scale limit, panels (a) and (b) show the temperature $\widetilde{T}$ plot and the progress variable source $\overline{\omega_C}$ plot, respectively. For comparison, panels (c) and (d) correspond to the results with some larger filter length $\sigma$ and $D_t$, as in a coarser LES spatial resolution setup. We can see that at the fine scales, the $\widetilde{T}$ and $\overline{\omega_C}$ distributions are narrower, and preserve sharp gradients and relatively high magnitudes. Differently, the coarser grid results exhibit attenuated peak values and broader distribution because of larger sized filtering and subgrid diffusion as well. These trends demonstrate that the subgrid diffusion, filter size and flame wrinkling effects are integrated into the tabulation database in a physically reasonable manner.

\section{Model applications}\addvspace{10pt}

In the following, two different jet flames will be tested using the newly developed FTTM-ST model, in together with the flamelet/progress variable (FPV)~\cite{PIERCE_MOIN_2004} model and FTFM-C~\cite{He13082025}, within the same LES solver with the identical computational mesh, inflow conditions, boundary conditions and other numerical settings clarified thereafter. Meanwhile, some other combustion model results in the existing literature 
are also presented for comparison.

For all the test cases, the inlet turbulence is generated using a synthetic turbulence inflow generator based on the random-spot method~\cite{Kornev} encoded in OpenFOAM-8 with the inlet Reynolds stresses prescribed from experimental measurements. The GRI-Mech 3.0 mechanism~\cite{gri30} is adopted as the combustion kinetics. In LES, the filter length $\sigma$ is determined directly from the local mesh size and the subgrid turbulent diffusivity $D_t$ is obtained dynamically from the $k$-equation subgrid-scale model. The Pressure-Implicit Method for Pressure-Linked Equations (PIMPLE) algorithm is also employed with the maximum Courant number limited to $0.5$ to ensure the numerical stability. Especially, it is worth noting that in model implementation, the value range of table input parameters needs to fit the numerical setup specification. For instance, the tabulated filter length should be able to cover the variation of the domain mesh size. For the following two cases, it is enough to set the filter length $\sigma$ from $10^{-5.0}\ \mathrm{m}$ to $10^{-2.2}\ \mathrm{m}$, and the subfilter diffusivity $D_t$ from $10^{-6.2}\ \mathrm{m^2s^{-1}}$ to $10^{-2.2}\ \mathrm{m^2s^{-1}}$. The obtained four-dimensional look-up table dimension is then $141$ (in $Z$) $\times 106$ (in $C$) $\times 15$ (in $\sigma$) and $\times 20$ (in $D_t$).

\subsection{Sydney swirl flame SMH1} \addvspace{10pt}

\begin{figure}[!h]
  \centering
  \subfigure[]{
    \centering
    \includegraphics[width=0.7\linewidth]{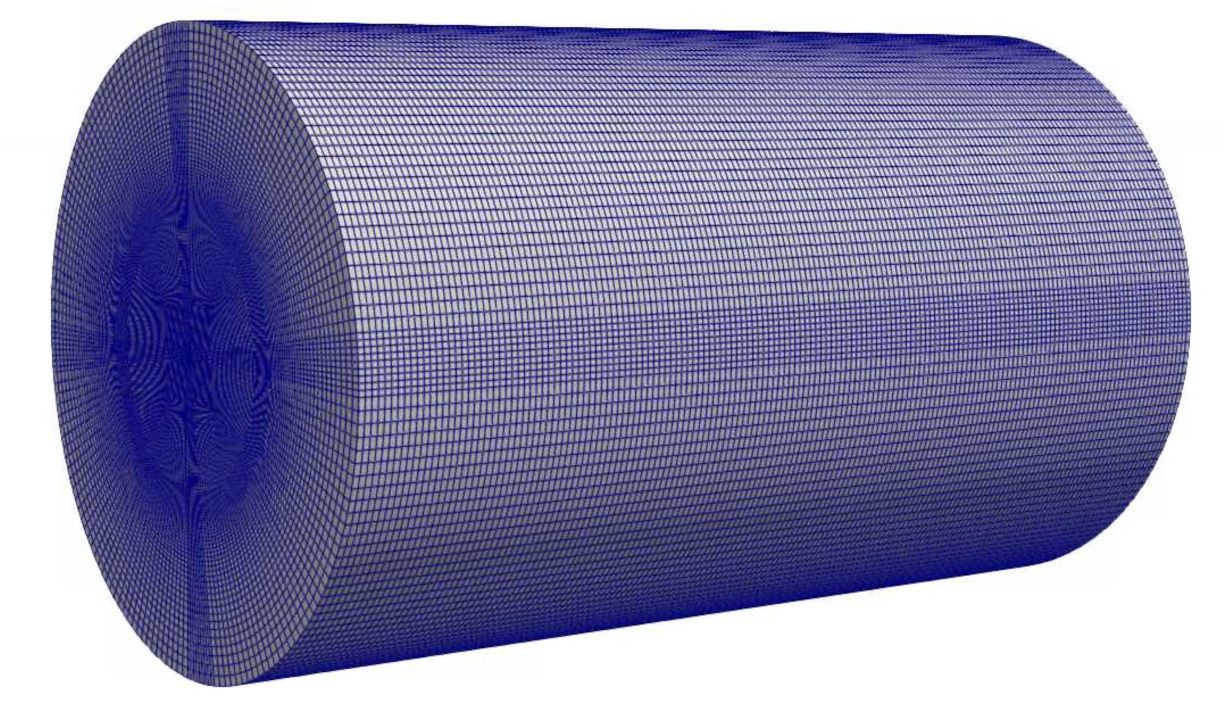}}
  \vspace{-0.5mm}%
  \subfigure[]{
    \centering
    \includegraphics[width=0.7\linewidth]{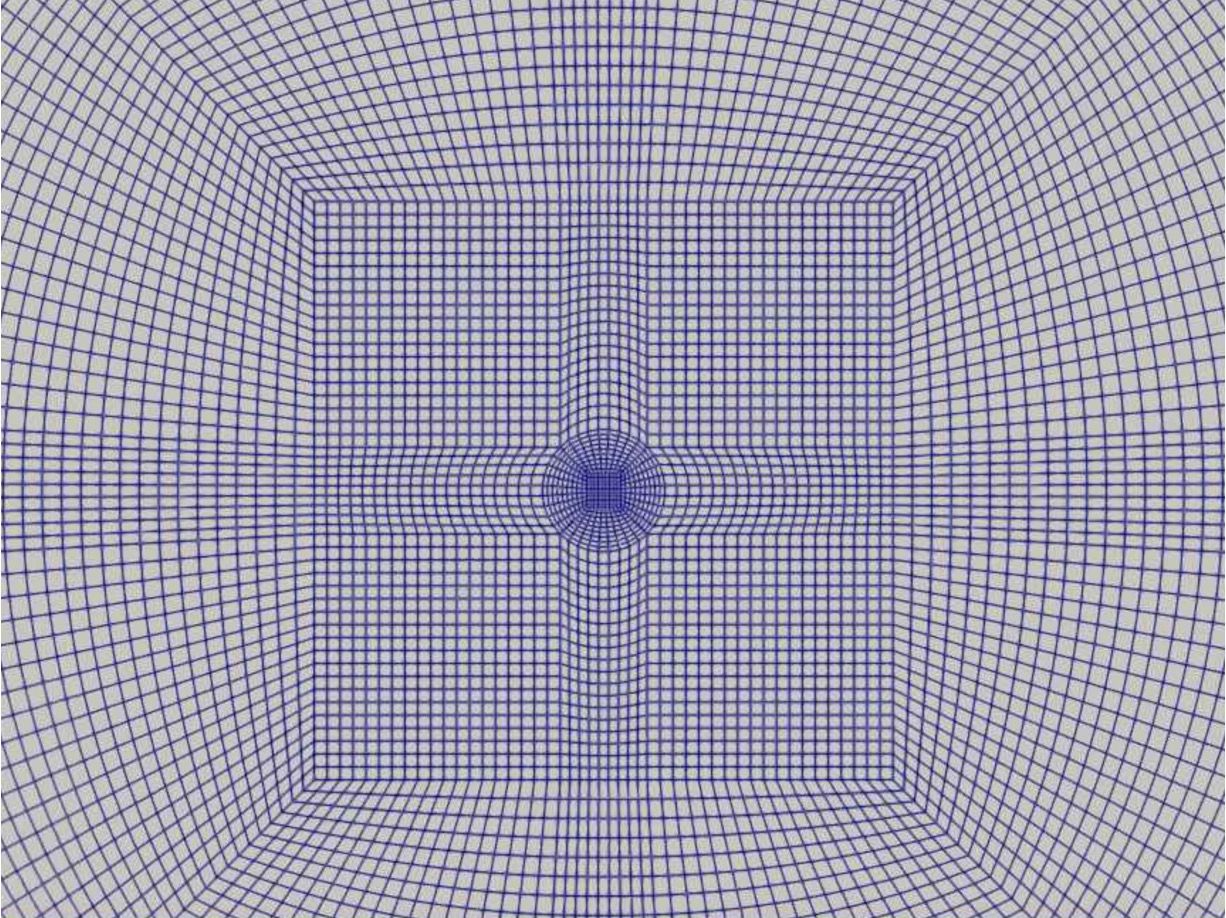}}
  \caption{(a) Computational domain of the Sydney swirl burner with radius $60\  \mathrm{mm}$ and length $200\  \mathrm{mm}$; (b) Inlet meshing details with different structured zonal partitions.}
  \label{fig:Sydinletconfig}
  \vspace{-5mm}
\end{figure}

\begin{figure*}[ht!]
  \centering
\subfigure[]{
    \raisebox{6mm}{
    \includegraphics[width=0.2\linewidth]{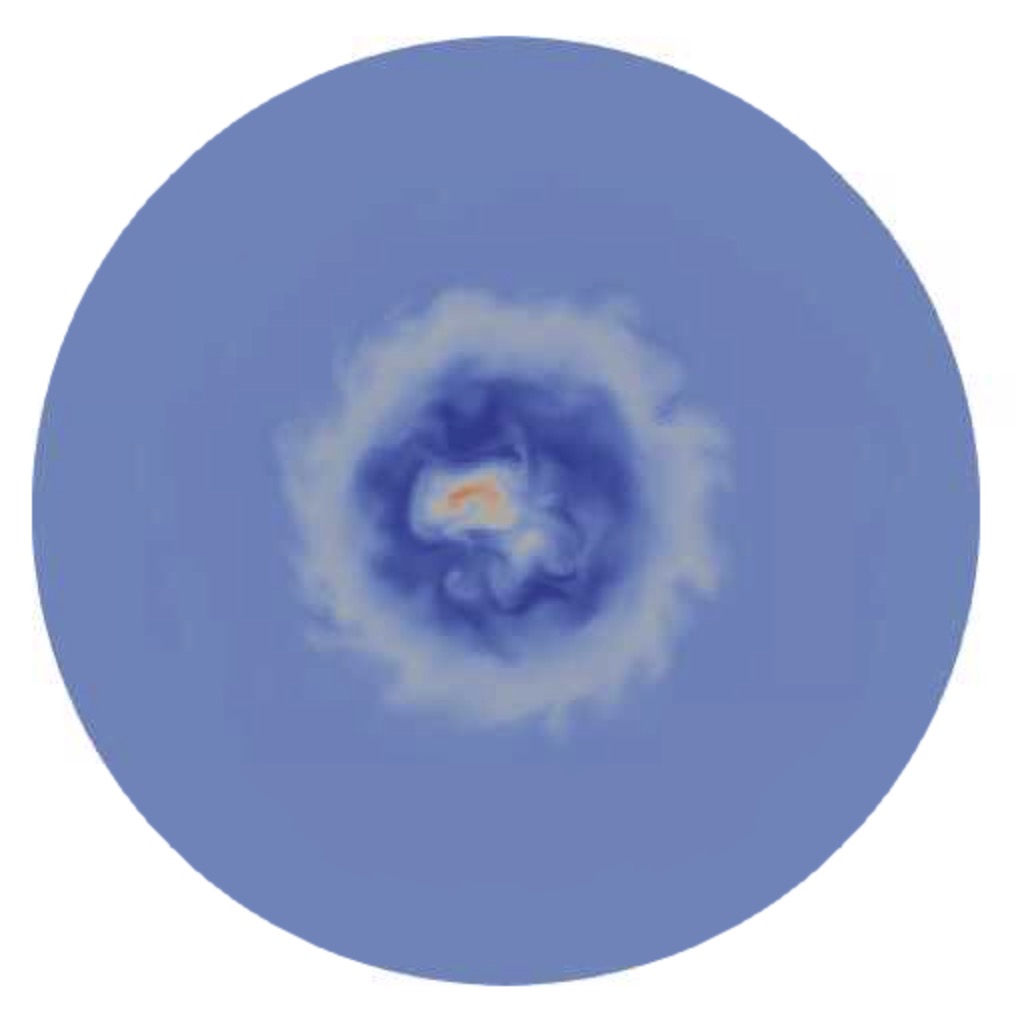}}}
\subfigure[]{
    \raisebox{6mm}{
    \includegraphics[width=0.2\linewidth]{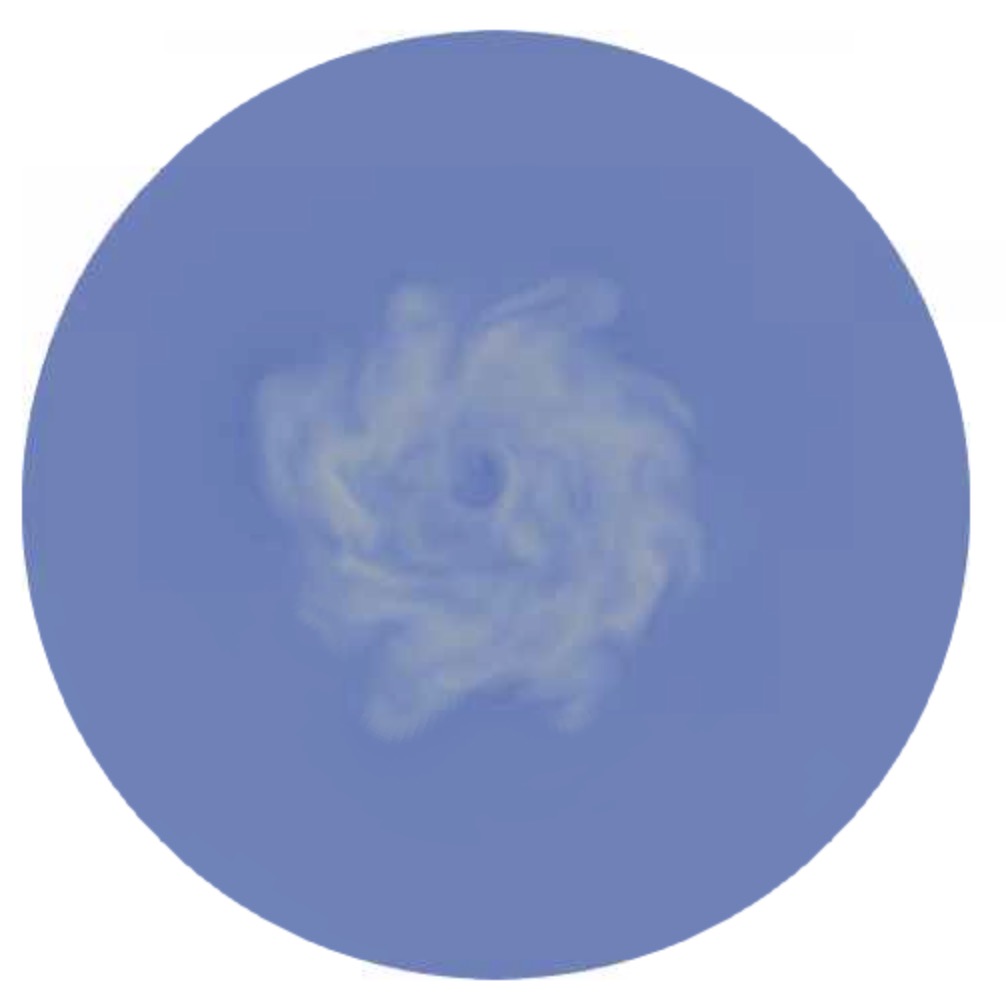}}}
\subfigure[]{
    \includegraphics[width=0.49\linewidth]{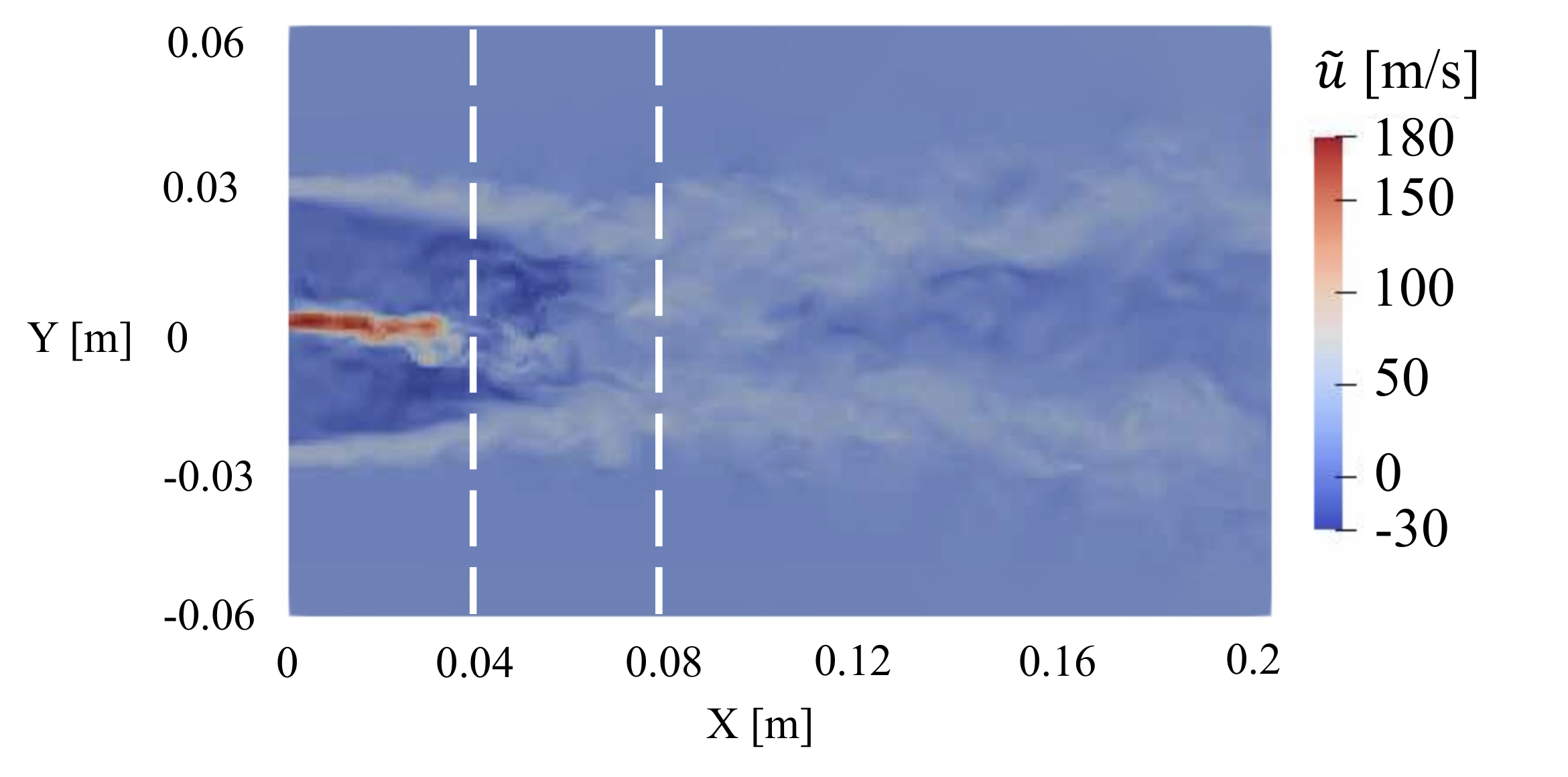}}

\subfigure[]{
    \raisebox{5.5mm}{\hspace*{3mm}
    \includegraphics[width=0.195\linewidth]{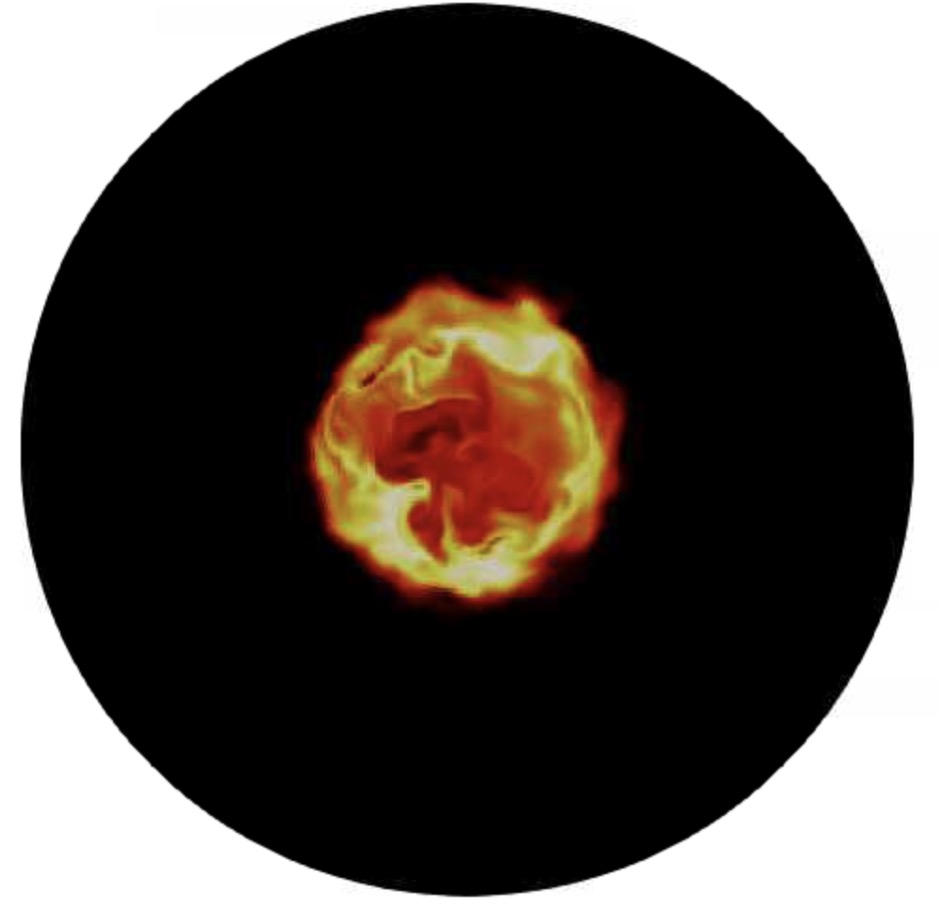}}}
\subfigure[]{
    \raisebox{5.5mm}{\hspace*{1mm}
    \includegraphics[width=0.195\linewidth]{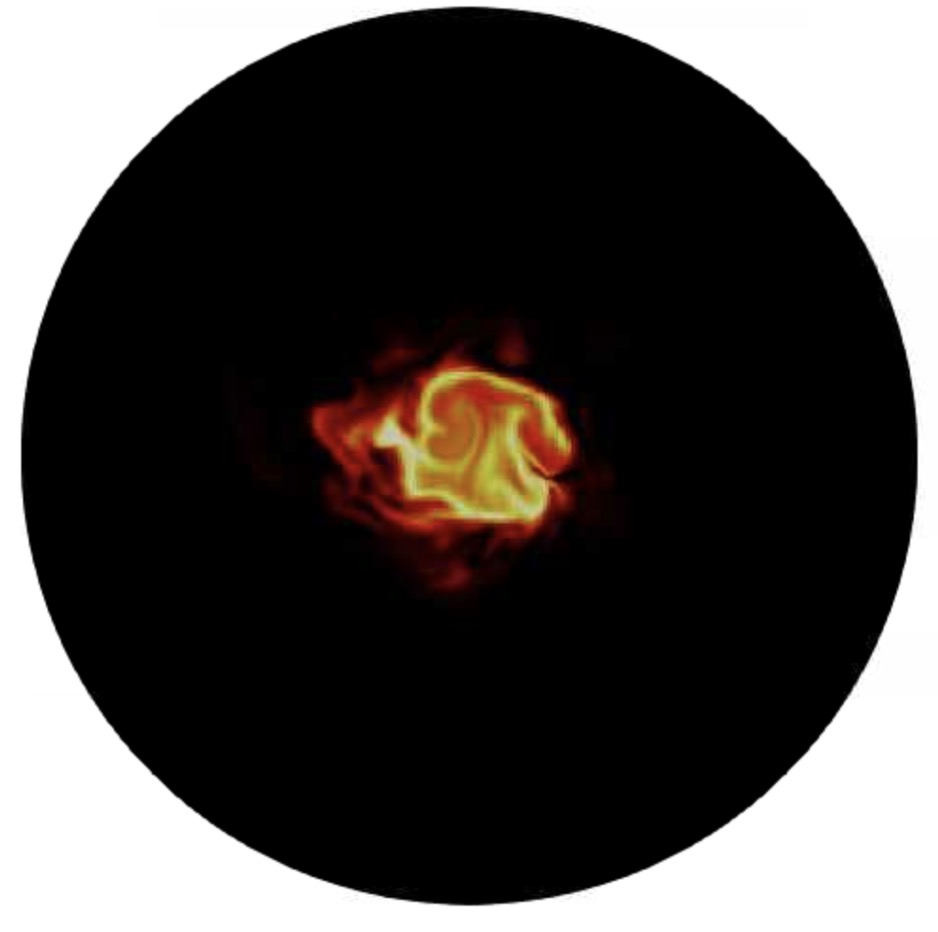}}}
\subfigure[]{\hspace*{0.5mm}
    \includegraphics[width=0.51\linewidth]{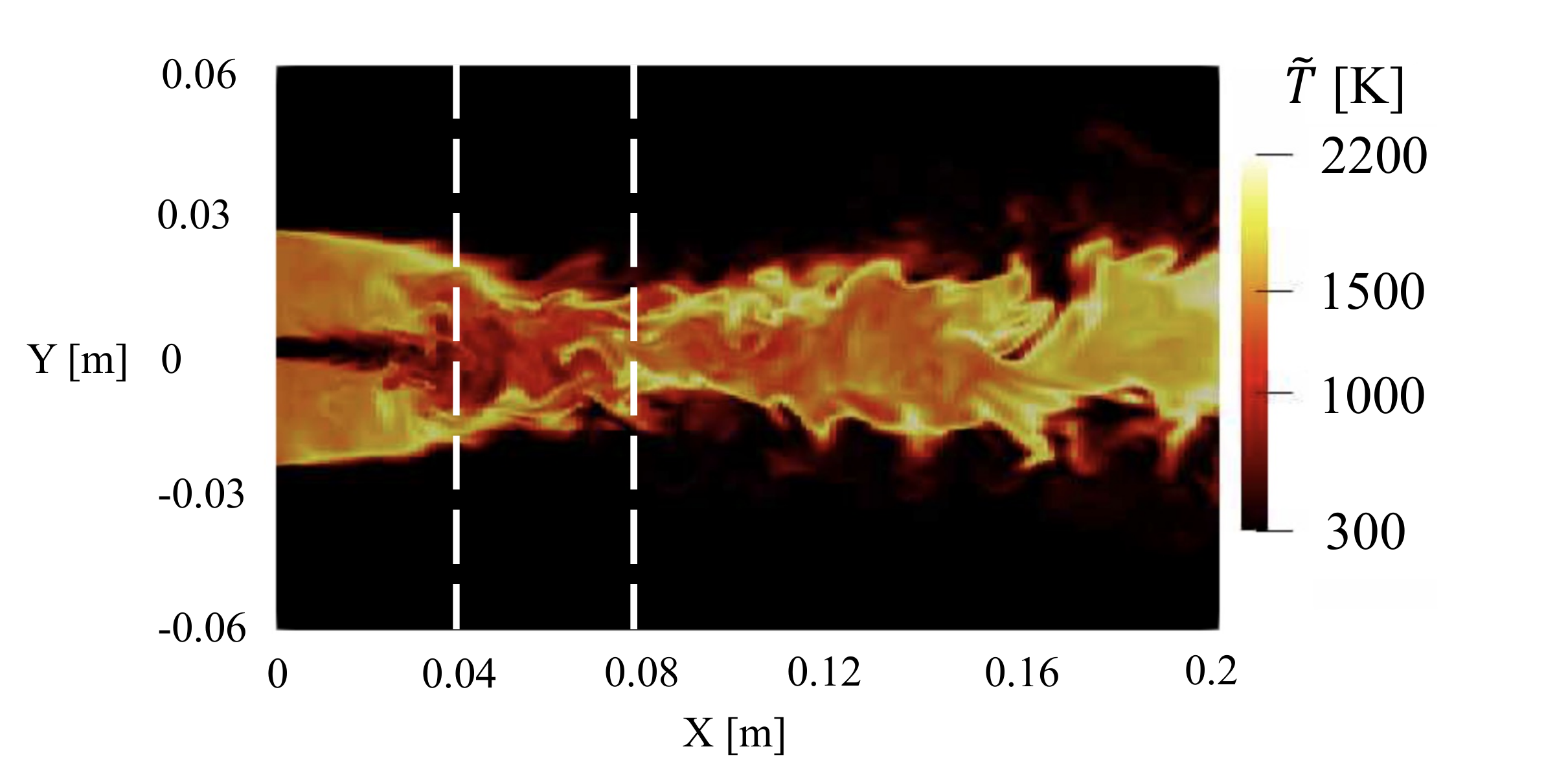}}
\caption{Instantaneous flow and flame structures of the Sydney swirl flame, including the transverse contour of axial velocity at: (a) $x = 40$ mm, (b) $x=80$ mm; (c) longitudinal contour of the axial velocity; transverse contours of temperature at: (d) $x=40$ mm, (e) $x=80$ mm; (f) longitudinal contour of the temperature.}
\label{fig:SwirlContours}
\end{figure*}

\begin{figure*}[h!]
\centering\vspace{-1mm}
\includegraphics[width=0.99\textwidth]{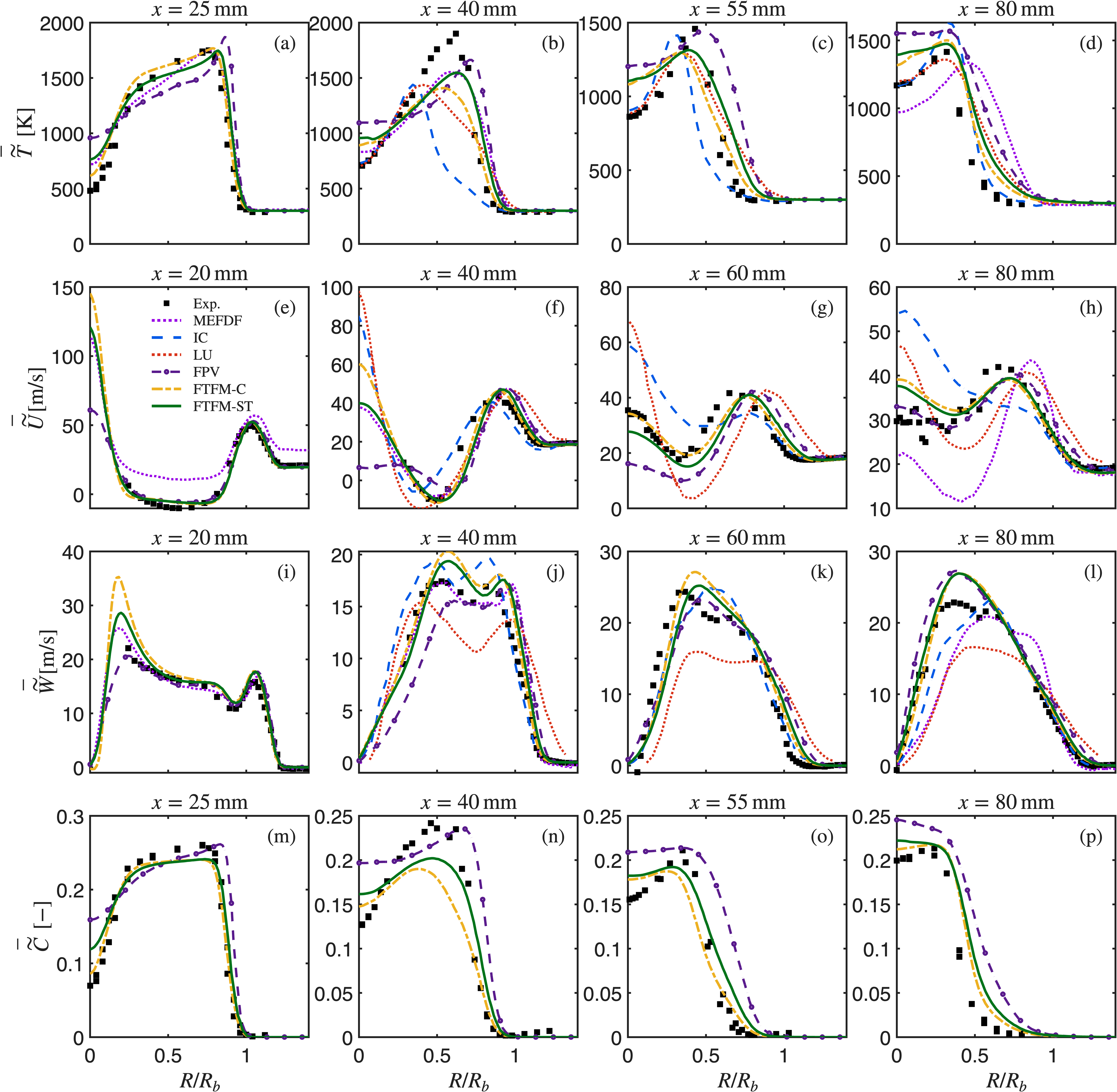}
\caption{Comparison of different model predictions of time-averaged quantities at distinct locations. For the temperature and progress variables profiles, columns from left to right correspond to the axial locations at $x=25$, $40$, $55$ and $80~\mathrm{mm}$, respectively. For the axial velocity component and swirling velocity component profiles, columns from left to right correspond to $x=20$, $40$, $60$ and $80~\mathrm{mm}$, respectively. The progress variable is defined as  $\overline{\widetilde{C}}=\overline{\widetilde{Y}}_{\text{H2O}}+\overline{\widetilde{Y}}_{\text{CO2}}+\overline{\widetilde{Y}}_{\text{CO}}$.}
\label{fig3:swirlplots}\vspace{-1mm}
\end{figure*}

The fuel used in the Sydney swirl flame SMH1 ~\cite{swirl_flows_database} is a methane/hydrogen mixture with a $1:1$ volumetric ratio. The fuel jet with a radius $R_b=3.6\ \mathrm{mm}$ is enclosed by a bluff-body with a diameter $D_0=50\ \mathrm{mm}$, surrounded by a swirling annular airflow with a radius $R_s = 5\ \mathrm{mm}$. Because of the rich physics from the interaction between the swirl flow and flame, the Sydney swirl flame is considered as a representative modeling benchmark. The bulk velocity of the fuel jet $U_{\mathrm{jet}}=140.8\ \mathrm{m/s}$. The axial and azimuthal mean velocity of the swirling flow are $U_s=42.8\ \mathrm{m/s}$ and $W_s=13.8\ \mathrm{m/s}$, respectively. The external coflow enters the wind tunnel with a square cross-section of $130^2\ \mathrm{mm}^2$ . The computational domain and inlet configuration are presented in Fig.~\ref{fig:Sydinletconfig}. The mesh consists of about $2.5$ million partitioned hexahedral cells. Figure~\ref{fig:Sydinletconfig} (b) particularly shows the structured zonal partition to sufficiently resolve the inlet structure. For the sake of brevity, other numerical setup details can be referred to  Refs.~\cite{He13082025,MASRI2000123}.

The instantaneous contours of the axial velocity component and temperature in both transverse and longitudinal crosscuts are presented in Fig.~\ref{fig:SwirlContours}. At $x=40$ mm, the transverse velocity field in panel (a) exhibits a pronounced recirculation zone with negative $\tilde{u}$ generated by the strong swirling motion. The recirculation continuously transports high-temperature burned products towards the burner exit, contributing to the flame stabilization. The high temperature core in panel (d) is anchored inside this recirculation zone. The bright region with higher temperature implies the main reaction there. In the peripheral region outside the recirculation zone $\tilde{u}$ is relatively large because of contribution of both mass and momentum from the swirling annular flow. Further downstream at $x=80$ mm, the central recirculation region gradually disappears because of the decay of swirl intensity and enhanced turbulent mixing. Meanwhile, the high temperature zone shrinks back to the central part because of the contraction of the fuel distribution there without centrifugal transport. The longitudinal contours in panels (c) and (f) further show that the flow undergoes vortex breakdown downstream, which is also observed by other works~\cite{SteinPCI2007,kempf2008}. Beyond $x=80$ mm, the jet flow evolves and approaches a fully developed state. These observations suggest that turbulence-induced flame wrinkling is expected to be most significant in the upstream recirculation region and gradually weaken farther downstream, which will be quantified by predictions from the proposed FTFM-ST and other models as well.

Figure~\ref{fig3:swirlplots} collectively shows the time-averaged predictions of the radial distributions of temperature $\overline{\widetilde{T}}$, axial velocity $\overline{\widetilde{U}}$, swirling velocity $\overline{\widetilde{W}}$, mixture fraction $\overline{\widetilde{Z}}$ and progress variable $\overline{\widetilde{C}}=\overline{\widetilde{Y}}_{\text{H2O}}+\overline{\widetilde{Y}}_{\text{CO2}}+\overline{\widetilde{Y}}_{\text{CO}}$ at different axial locations from different approaches, including the multi-environment filtered density function model (MEFDF)~\cite{zhang2013les}, laminar flamelet model results~\cite{kempf2008}, FPV~\cite{PIERCE_MOIN_2004}, FTFM-C~\cite{He13082025} (without filtered chemical source stretching), the present FTFM-ST with filtered chemical source stretching and the experimental data~\cite{swirl_flows_database}. In the upstream region, FTFM-C, FTFM-ST and FPV models can reasonably capture recirculation region, i.e. the negative $\tilde{U}$ part in panel (e), while no recirculation part can be observed from MEFDF. Moving downstream, mixing becomes intensified as the flow develops, leading to more pronounced discrepancies among these models.

Overall, FTFM-C and FTFM-ST, the same type of models, yield satisfactory accuracy in predicting the radial spreading near the flame edge, where sharp gradients and strong interactions between the reactants are present, whereas the other models (IC, LU, FPV and MEFDF) exhibit more deviations to various extents. We address here that the distinction between FTFM-C and FTFM-ST emerges particularly at $x=40\ \mathrm{mm}$ for temperature and progress variable profiles, where FTFM-ST has a higher peak temperature together with a broader reaction zone, which match better the experimental points. Such an improvement is from the flame wrinkling correction in Eq.~\eqref{eq:stretch1}. Further downstream, the differences between FTFM-ST and FTFM-C gradually diminish, which is consistent with the weakening of the recirculation and turbulence levels shown in Fig.~\ref{fig:SwirlContours}. Once turbulence decays downstream, the stretching correction because of $D_t$ is thus weaker. In general, these observations are consistent with the discussion in Section~\ref{FWdiscussion}. 

\begin{figure*}[h!]
\centering\vspace{-1mm}
\includegraphics[width=0.99\textwidth]{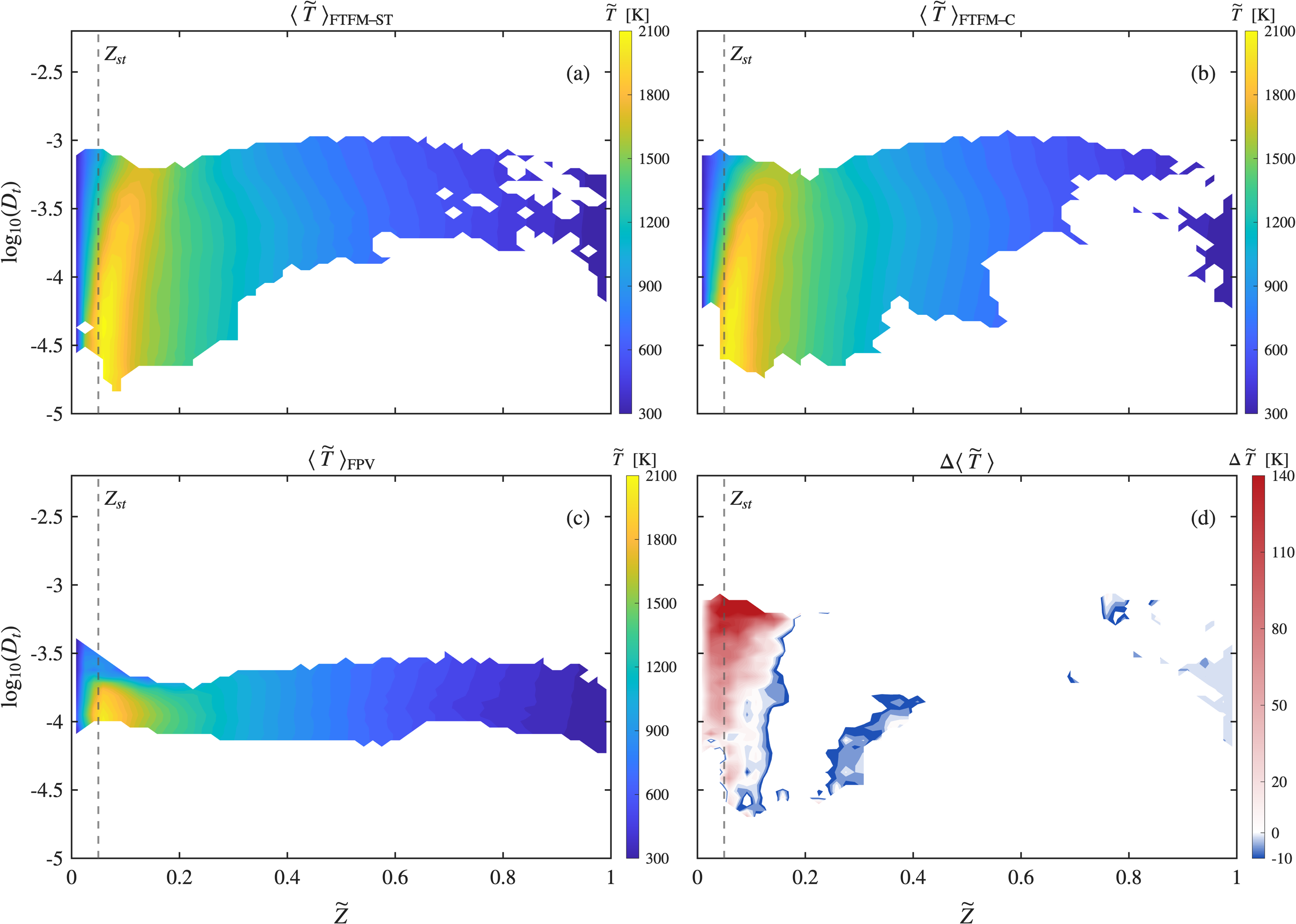}
\caption{Predicted mean temperature conditioned on given $\widetilde{Z}$ and $\log_{10}(D_t)$, from: (a) FTFM-ST, (b) FTFM-C, and (c) FPV. Panel (d) shows the difference between the two FTFM-C and FTFM-ST, defined as $\Delta\langle\widetilde{T}\rangle\equiv\langle\widetilde{T}\rangle_{\mathrm{FTFM\!-\!ST}}-\langle\widetilde{T}\rangle_{\mathrm{FTFM\!-\!C}}$. The vertical dashed lines mark the stoichiometric mixture fraction $Z_{st}=0.0495$.} \label{fig:Cond}\vspace{-1mm}\end{figure*}

To further explore the model difference, Fig.~\ref{fig:Cond} presents the temperature conditionally averaged with respect to the resolved mixture fraction $\widetilde{Z}$ and turbulent diffusivity $D_t$. From the post processing of the simulation data, panels (a)-(d) show the results from FTFM-ST, FTFM-C, FPV, and difference between FTFM-C and FTFM-ST, respectively. The FTFM-C and FTFM-ST plots exhibit similar patterns, while the FPV distribution is much narrower along the $D_t$ dimension. The high temperature regions for all the models are around the stoichiometric mixture fraction $Z_{st}$ state. It is also interesting to see that in these regions the temperature value decreases if $D_t$ increases, which physically can be understood from the fact that at the high temperature the enhanced fluid viscosity will suppress the turbulence intensity and $D_t$ value as well. The panel (d) show that generally the temperature predicted from FTFM-ST is higher than that from FTFM-C, i.e. the source-stretching correction in FTFM-ST leads to a systematic temperature increment. Such temperature difference exceeds approximately $100\ \mathrm{K}$, and becomes larger as $D_t$ increases. This behavior is consistent with the one-dimensional flamelet results in Fig.~\ref{Fig:flameletSol1}.

\subsection{Delft III Piloted Natural Gas Flame} \addvspace{10pt}

\begin{figure}[!h]
\centering
\subfigure[]{
    \includegraphics[width=0.94\linewidth]{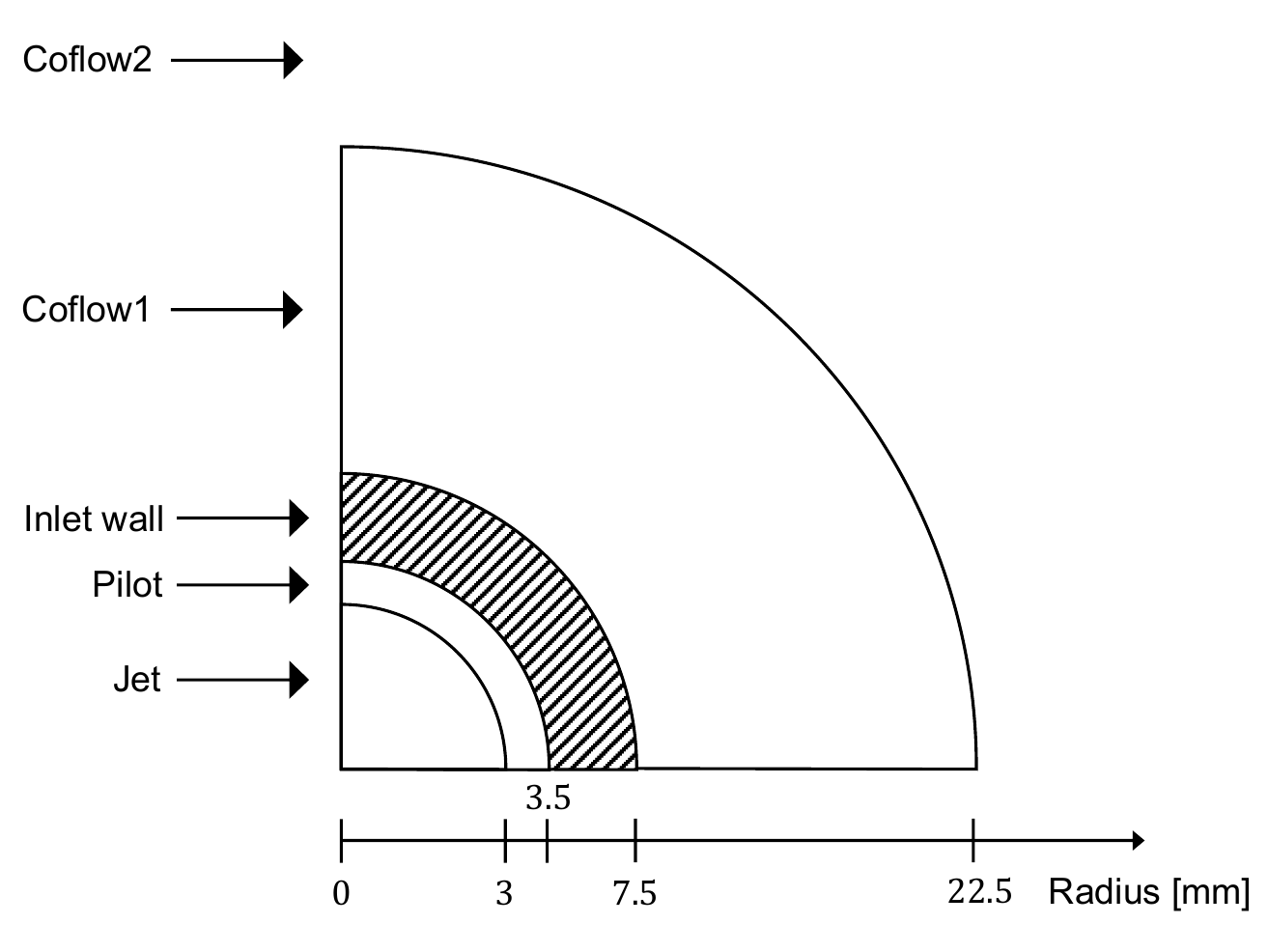}}
\vspace{-0.5mm}
\subfigure[]{
    \includegraphics[width=0.54\linewidth]{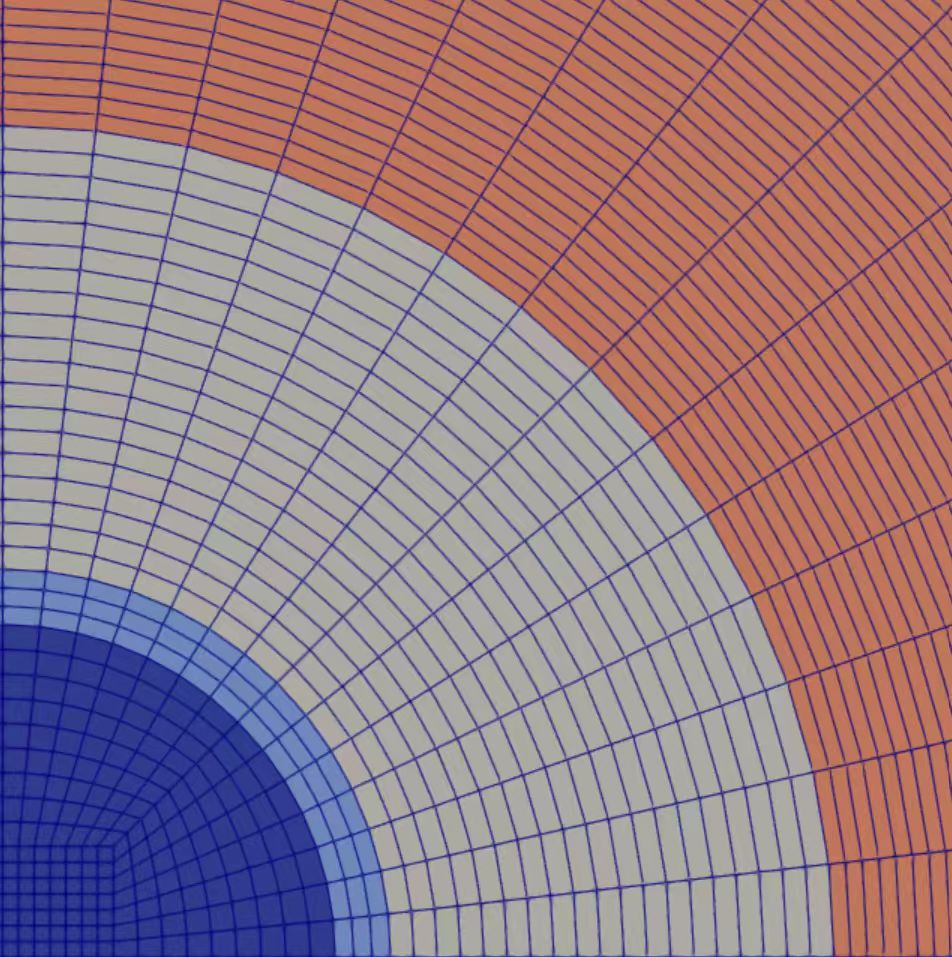}}
\caption{(a) Inlet configuration of the Delft piloted jet burner; (b) Computational mesh of inlet (local). Different colors correspond to different zones in panel (a).}
\label{fig4:inletconfig}
\end{figure}

For the Delft III piloted natural gas flame~\cite{delft3_database}, the fuel is the Dutch natural gas, which is numerically treated as $85.3\%\ \text{CH}_4$ and $14.7\%\ \text{N}_2$ by volume ~\cite{ayache2012cmcles,zakyani2022delft}. As shown in the nozzle structure in Fig.~\ref{fig4:inletconfig} (a), to stabilize the flame, the central injected fuel is surrounded by twelve pilot flames at radius $R=3.5\ \mathrm{mm}$, each with a $0.5\ \mathrm{mm}$ orifice diameter. For simplicity, these pilots are modeled as an annular ring with an equivalent width of $0.5\ \mathrm{mm}$ around the central jet, maintaining the same total mass flow rate. More detailed numerical setups can be referred to Ref.~\cite{ayache2012cmcles}. Totally there are about $1.6$ million computational cells, with local refinement around the jet center, as shown in mesh partition Fig.~\ref{fig4:inletconfig} (b).

\begin{figure}[!h]
\centering
\subfigure[]{
    \centering
    \includegraphics[width=0.995\linewidth]{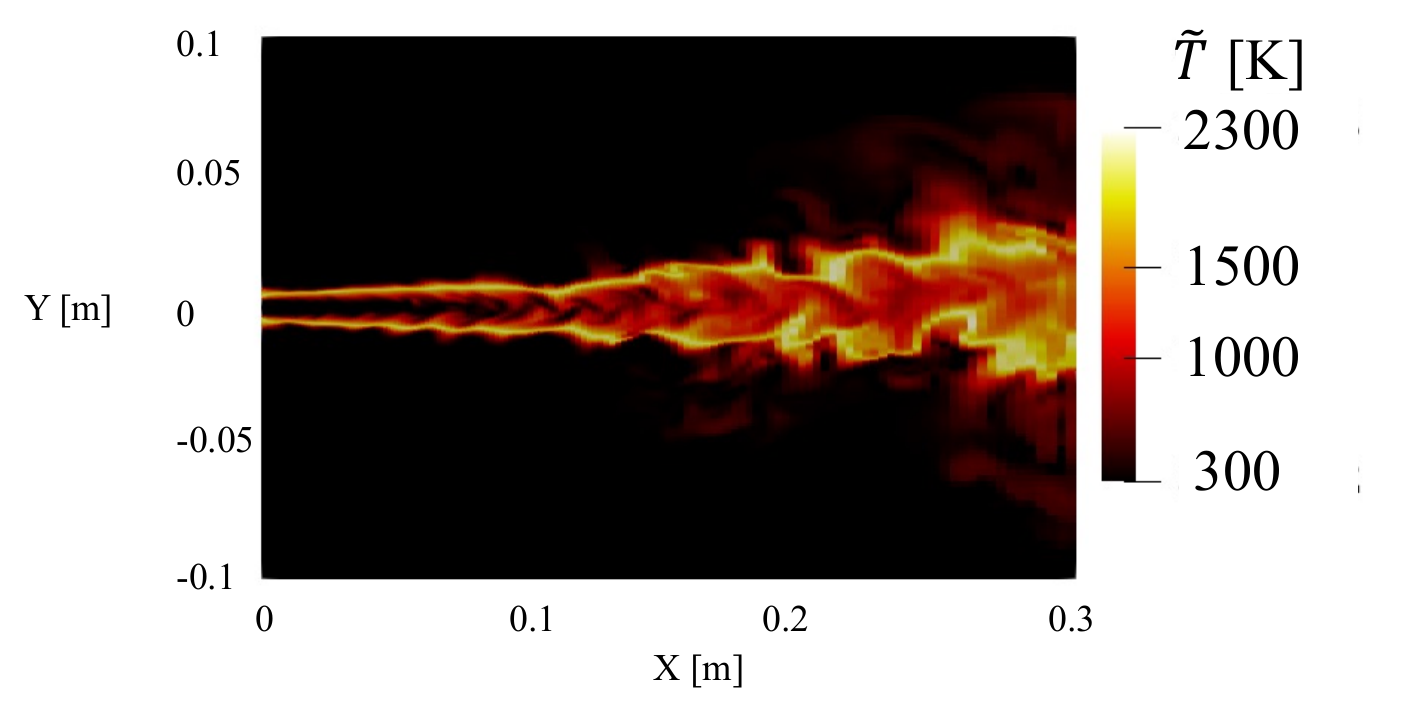}}
\vspace{-0.5mm}
\subfigure[]{
    \centering
    \includegraphics[width=0.995\linewidth]{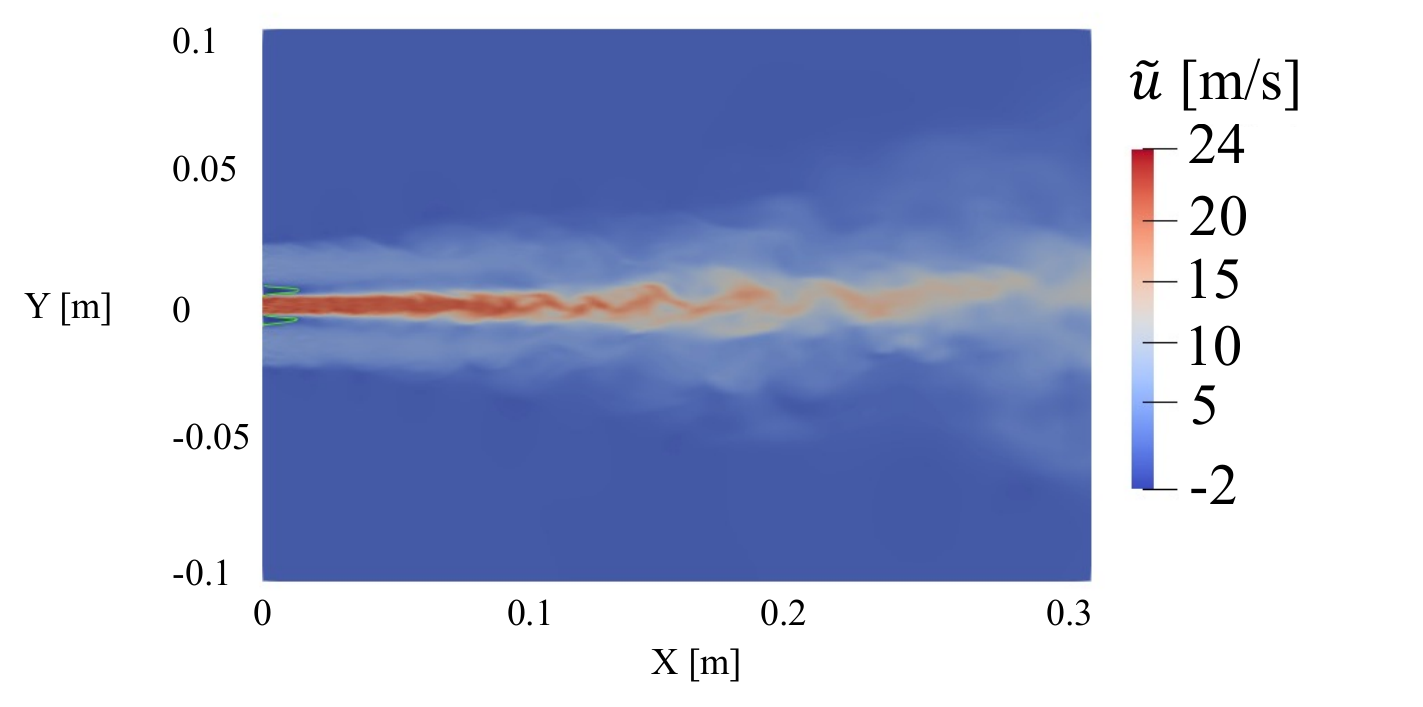}}
\caption{Instantaneous contours (longitudinal cut) for: (a) temperature; (b) axial velocity. The green zero-isoline close to the jet inlet marks the recirculation zone.}
\label{fig5:insplots}
\end{figure}

The instantaneous contours of the temperature $\tilde{T}$ and axial velocity component $\tilde{u}$ on the longitudinal plane are presented in Fig.~\ref{fig5:insplots}. At the inlet, the flame is around the central fuel jet. Moving downstream, the turbulent flame zone becomes thicker and cold core region gradually disappears. In addition to the pilot flame, a small recirculation zone, shown as the region inside the green zero-isoline of $\tilde{u}$ in Fig.~\ref{fig5:insplots} (b), develops near the rim of the inlet wall, which also contributes to flame stabilization. The wavy structure of the stream mixing interface in both panel (a) and (b) indicates the Kelvin-Helmholtz instability induced by density stratification.

\begin{figure*}[h!]
\centering\vspace{-1mm}
\includegraphics[width=0.99\textwidth]{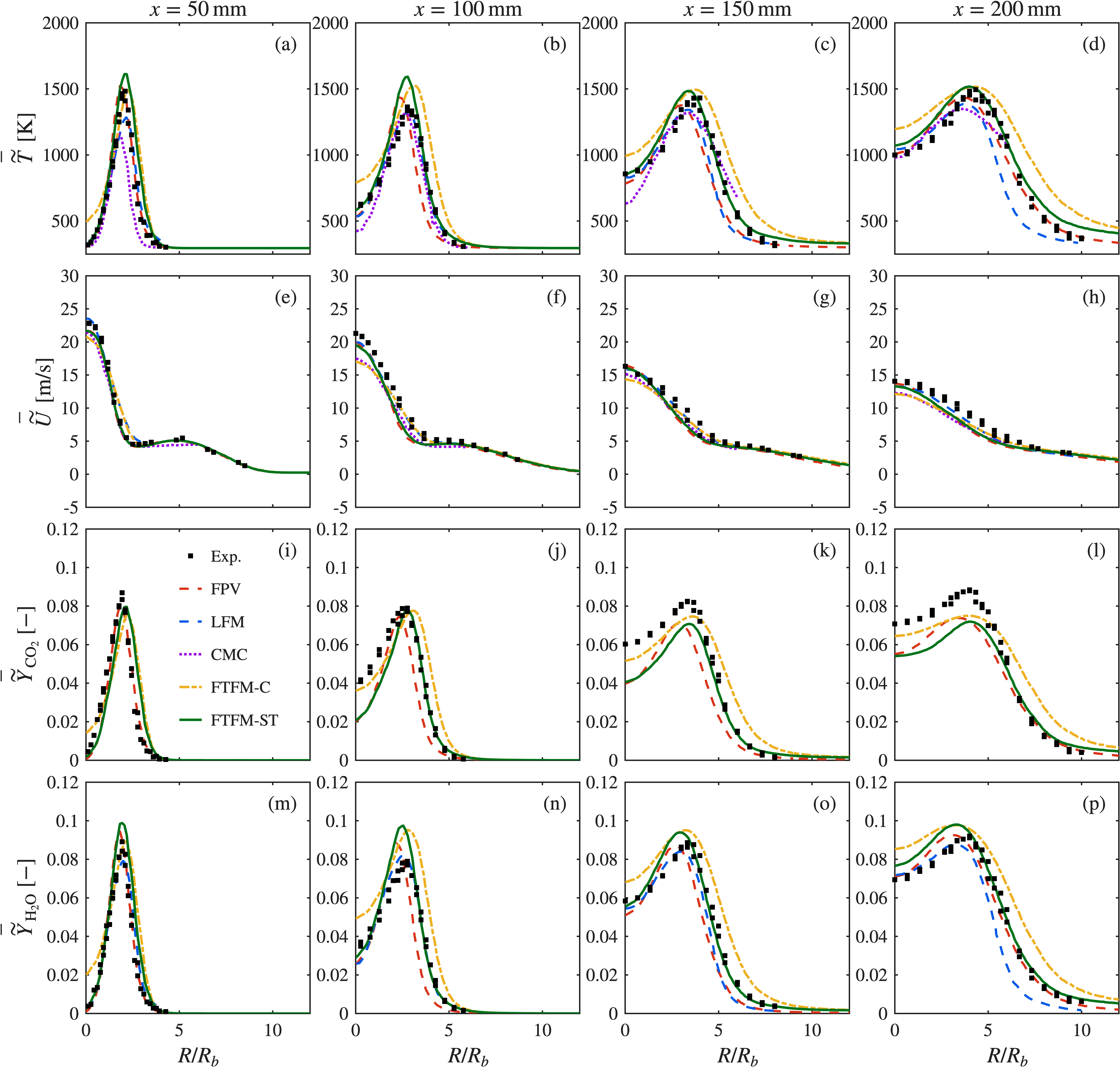}
\caption{Predictions of the mean field quantities from different models at distinct axial locations $x=50$, $100$, $150$ and $200~\mathrm{mm}$, including: temperature $\overline{\widetilde{T}}$ in panel $a$-$d$; axial velocity $\overline{\widetilde{U}}$ in panel $e$-$h$; $\text{CO}_2$ mass fraction $\overline{\widetilde{Y}}_{\text{CO}_2}$ in panel $i$-$l$; $\text{H}_2\text{O}$ mass fraction $\overline{\widetilde{Y}}_{\text{H2O}}$ in panel $m$-$p$.}
\label{fig:6subplots}\vspace{-1mm}
\end{figure*}

Figure~\ref{fig:6subplots} presents the time-averaged radial profiles of temperature, axial velocity, and species mass fractions at distinct axial locations $x=50, 100, 150, 200\ \mathrm{mm}$, predicted by FPV, laminar flamelet model (LFM)~\cite{LFM2005}, conditional moment closure method (CMC)~\cite{ayache2012cmcles}, FTFM-C and the present FTFM-ST model, against experimental measurements. Overall, all models reproduce the mean velocity profiles properly. FTFM-ST yields more accurate predictions of the temperature and species, particularly in terms of the peak locations and radial spreading of the flame front. It is also worth noting that near the flame edge, FTFM-ST generally provides improved predictions. For the $Y_\mathrm{CO_2}$ profiles, a bit more discrepancies appear, regardless of the choice of the progress variable and combustion models, which needs to be further tested.

\begin{figure*}[htbp!]
\centering
\includegraphics[width=0.9\linewidth]{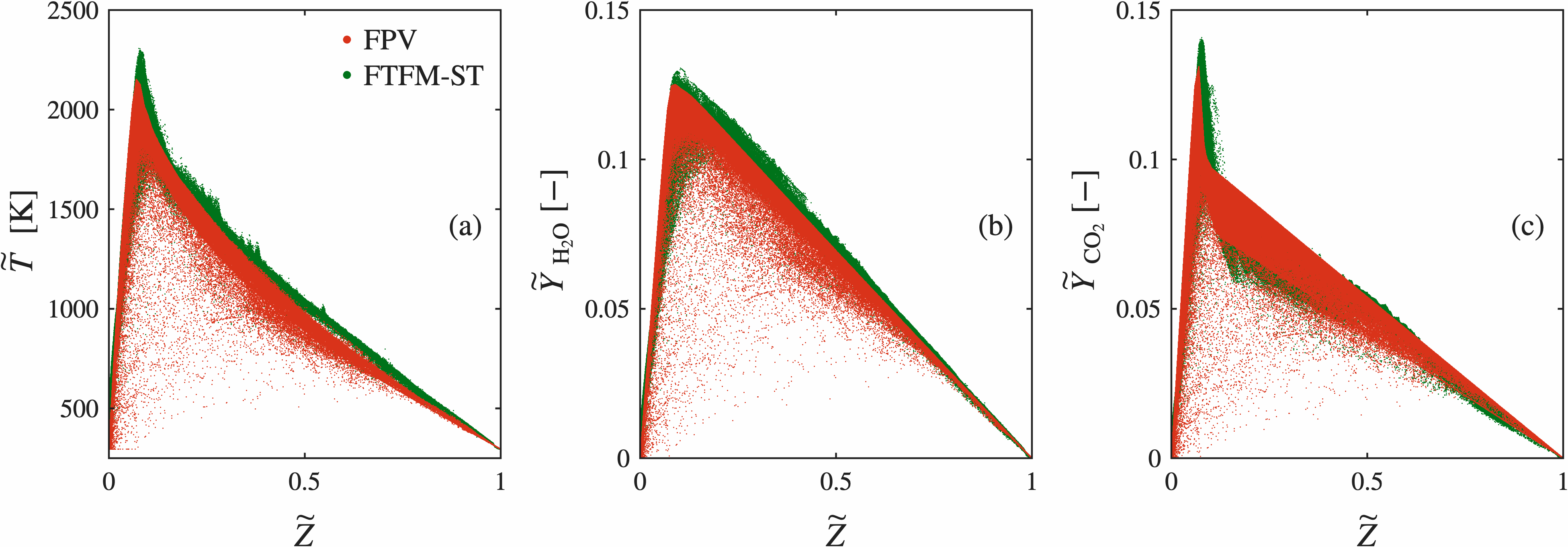}
\caption{Scatter plots of the Delft III flame data predicted from FPV and FTFM-ST: (a) $\tilde{Z}$--$\tilde{T}$; (b) $\tilde{Z}$--$\tilde{Y}_{\mathrm{H_2O}}$; (c) $\tilde{Z}$--$\tilde{Y}_{\mathrm{CO_2}}$.}
\label{fig10:mappingplots}
\end{figure*}

More understanding of the model performance between FPV and FTFM-ST can be gained from Fig.~\ref{fig10:mappingplots}, showing the $\tilde{Z}$ mapped relations, including $\tilde{Z}-\tilde{T}$, $\tilde{Z}-\tilde{Y}_{\mathrm{H_2O}}$ and $\tilde{Z}-\tilde{Y}_{\mathrm{CO_2}}$. It can be seen that scattering from FTFM-ST is more concentrated. Since flame appears basically
around the stoichiometric surfaces, both models demonstrate identical peak locations. Relatively, the peaks values from FTFM-ST are higher.

\section{Conclusion} \addvspace{10pt}
Regarding the recently developed non-premixed filtered turbulent flame model (FTFM), its unique modeling advantages can be clarified from the following two important aspects. First, in the framework of large eddy simulations (LES), the table look-up quantities are directly the filtered ones, for instance the turbulent scalar dissipation rate and filtered mixture fraction. Therefore, there is no need to introduce assumed probability density functions (PDFs) to model these numerically available quantities. Second, for the notorious nonlinear chemical sources, the one-to-one correspondence between the filtered thermochemical quantities and the filtered chemical sources is determined by inversely solving the filtered counterflow flame equations. 

In the present work, FTFM is further updated to FTFM-ST with chemical source stretching (ST), on the basis of the analytical formulation of the flame wrinkling correction on the filtered chemical source profile. Specifically, the profile peak needs to be retained, whereas the integrated consumption rate is enhanced by a prescribed wrinkling factor because of the stretched flame width. In model validation, the model predictions of both the Sydney swirl flame and the Delft III flame can be reasonably improved, in comparison with conventional models such as LFM, FPV and CMC. The model performance has been explored in details from the structure of the tabulated database. Overall, FTFM-ST provides a turbulent combustion physics founded modeling strategy for non-premixed turbulent combustion simulations.

\section*{CRediT authorship contribution statement} \addvspace{10pt}
{\bf Haoyu Lu}: Writing - original draft, Methodology, Formal analysis, Investigation, Software, Data curation, Visualization, Validation. {\bf Junyi He}: Methodology, Formal analysis, Investigation, Software. {\bf Lipo Wang}: Writing - review \& editing, Supervision, Project administration, Methodology, Investigation, Funding acquisition, Formal analysis, Conceptualization.

\section*{Declaration of competing interest} \addvspace{10pt}

The authors declare that they have no known competing financial interests or personal relationships that could have appeared to influence the work reported in this paper.

\section*{Acknowledgments} \addvspace{10pt}

The computational support from the Centre for High-Performance Computing at
Shanghai Jiao Tong University ($\pi$) is gratefully appreciated.


 \newpage


\bibliographystyle{cnf-num}
\bibliography{cnf-refs}

\end{document}